\documentstyle[aps]{revtex}
\begin{document}
\draft
\title{Static charged fluid spheres in general relativity}
\author{B.V.Ivanov}
\address{Institute for Nuclear Research and Nuclear Energy,\\
Tzarigradsko Shausse 72, Sofia 1784, Bulgaria}
\maketitle

\begin{abstract}
Interior perfect fluid solutions for the Reissner-Nordstr\"om metric are
studied on the basis of a new classification scheme. It specifies which two
of the characteristics of the fluid are given functions and accordingly
picks up one of the three main field equations, the other two being
universal. General formulae are found for charged de Sitter solutions, the
case of constant energy component of the energy-momentum tensor, the case of
known pressure (including charged dust) and the case of linear equation of
state. Explicit new global solutions, mainly in elementary functions, are
given as illustrations. Known solutions are briefly reviewed and corrected.
\end{abstract}


\section{Introduction}

The unique exterior metric for a spherically symmetric charged distribution
of matter is the Reissner-Nordstr\"om solution. Interior regular charged
perfect fluid solutions are far from unique and have been studied by
different authors. The case of vanishing pressure (charged dust (CD)) has
received considerable attention. The general solution, in which the fluid
density equals the norm of the invariant charge density, was presented in
curvature coordinates by Bonnor \cite{one}. The proof that this relation
characterizes regular CD in equilibrium, i.e. in the general static case,
was given later \cite{two,three}. In the spherically symmetric case another
proof was proposed in Ref.\cite{four}. Concrete CD solutions were studied in
these coordinates \cite{four,five}. The generalization of the incompressible
Schwarzschild sphere to the charged case with constant $T_0^0$ was also
undertaken in a CD environment \cite{six}. Charged dust, however, has been
investigated more frequently in isotropic coordinates, since these encompass
the entire static case and allow to search for interior solutions to the
more general Majumdar-Papapetrou electrovacuum fields \cite{seven,eight}. In
both coordinate systems there is a simple functional relation between $%
g_{00} $ and the electrostatic potential. In isotropic coordinates there is
one non-linear main equation \cite{seven,nine} which has been given several
spherical \cite{ten,eleven,twelve,thirteen} and spheroidal \cite
{eleven,twelve,fourteen} solutions. One of them coincides with the general
static conformally flat CD solution \cite{fifteen}. These CD clouds may be
realized in practice by a slight ionization of neutral hydrogen, although
the necessary equilibrium is rather delicate. They have a number of
interesting properties: their mass and radius may be arbitrary, very large
redshifts are attainable, their exteriors can be made arbitrarily near to
the exterior of an extreme charged black hole. In the spheroidal case the
average density can be arbitrarily large, while for any given mass the
surface area can be arbitrarily small. When the junction radius $r_0$
shrinks to zero, many of their characteristics remain finite and
non-trivial. One can even entertain the idea for a point-like classical
model of electron were it not for the unrealistic ratio of charge $e$ to
mass $m$ \cite{one}.

Recently, new static CD solutions were found, in particular with density
which is constant or is concentrated on thin shells \cite{sixteen,seventeen}%
. In the spherically symmetric case a relation has been established with
solutions of the Sine-Gordon and the $\lambda \phi ^4$ equations \cite
{eighteen}.

The necessary condition for a quadratic Weyl-type relation has been derived
also for perfect fluids with non-vanishing pressure \cite{nineteen,twenty}.
However, in this case many other dependencies between the electrostatic and
the gravitational potential are possible, even when combined with constant $%
T_0^0$ \cite{twentyone}.

The original Schwarzschild idea of constant density has been also tested in
the charged case for a perfect fluid \cite{twentytwo,twentythree,twentyfour}
or for imperfect fluid with two different pressures \cite{one}. An
electromagnetic mass model with vanishing density has been proposed in Ref. 
\cite{six}. Unfortunately, the fluid has negative pressure (tension).
Although the junction conditions do not require the vanishing of the density
at the boundary, this is true for gaseous spheres. A model with such type of
density was proposed both in the uncharged and the charged case \cite
{twentyfive,twentysix}.

Another idea about the electromagnetic origin of the electron mass maintains
that, due to vacuum polarization, its interior has the equation of state $%
\rho +p=0$, where $\rho $ is the density and $p$ is the pressure. This leads
to tension, easier junction conditions and realistic $e$ and $m$ \cite
{twentyseven,twentyeight,twentynine}. It can be combined with a Weyl-type
character of the field \cite{thirty}. The experimental evidence that the
electron's diameter is not larger than $10^{-16}$ cm, however, requires that
the classical models should contain regions of negative density \cite
{thirtyone,thirtytwo}. Probably an interior solution of the Kerr-Newman
metric is more adequate in this respect.

The presence of five unknown functions and just three essential field
equations allows one to specify the metric and solve for the fluid
characteristics \cite{thirtythree}. This is impossible in the uncharged
case. Another approach is to electrify some of the numerous uncharged
solutions. This has been done for one of the Kuchowicz solutions \cite
{thirtyfour} in Ref. \cite{thirtyfive}. Two other papers \cite
{thirtysix,thirtyseven} build upon the Wyman-Adler solution \cite
{thirtyeight,thirtynine}. Thus a charged solution is obtained, which has
approximately linear equation of state when $m/r_0$ is small. In Ref. \cite
{forty} a generalization of the Klein-Tolman (KT) solution \cite
{fortyone,fortytwo} was performed, but the resulting fluid does not possess
a linear equation of state. Recently, static uncharged stars with spatial
geometry depending on a parameter \cite{fortythree,fortyfour} have been
generalized to the charged case \cite{fortyfive}.

The purpose of the present paper is to present a new and simple
classification scheme for charged static spherically symmetric perfect fluid
solutions. The calculations in each case are pushed as further as possible
and general formulae are given in many instances. The known solutions are
reviewed and compartmentalized according to this scheme in order to
illustrate general ideas, without being exhaustive. New solutions are added
where appropriate. The intention has been to stick to the simplest cases and
remain in the realm of elementary if not algebraic functions. The junction
conditions and other requirements for physically realistic models are
discussed. The emphasis is, however, on the general picture, which appears
unexpectedly rich and simpler than in the uncharged case.

The metric of a static spherically symmetric spacetime in curvature
coordinates reads 
\begin{equation}
ds^2=e^\nu dt^2-e^\lambda dr^2-r^2d\Omega ^2,  \label{one}
\end{equation}
where $d\Omega ^2$ is the metric on the two-sphere and $\nu ,\lambda $
depend on $r$. The fluid and its gravitation are described by five functions
depending on the radius: $\lambda ,\nu ,\rho ,p$ and the charge function $q$
which measures the charge within radius $r$. There are only three essential
field equations, hence, two of the above characteristics must be given. We
shall classify solutions according to this feature. For example, $\left( \nu
,\lambda \right) $ is the case of given metric and the other three fluid
characteristics are found from the equations. This does not mean that
solutions are distributed among groups which do not overlap. Thus, $\left(
\rho ,q\right) $ is a completely general case - any solution, after $\rho $
and $q$ are known, may be put into this class. The essence is that $\rho $
and $q^2$ are given functions and there is control over them; they can be
chosen regular, positive and comparatively simple. Then the other three
functions are usually more complex and are not always physically realistic.

In Sec.II the Einstein-Maxwell equations are organized into three main and
two auxiliary ones. Two of the main equations are universal for all cases,
the third one varies from case to case. Cases with given $\nu $ have linear
first-order differential equations for $e^{-\lambda }$. Cases with given $%
\lambda $ have linear second-order equations for $e^{\nu /2}$ or non-linear
first order equations for $\rho +p$. These results hold also for fluids with
a linear equation of state. The difficulties in the cases $\left( \rho
,p\right) $ and $\left( \rho ,q\right) $ are discussed, which prevent their
analytical treatment. The junction conditions are given in general form and
a reasonable set of physical requirements is included.

In Sec.III the case $\left( \lambda ,\nu \right) $ is briefly discussed.
More attention is paid to the condition $T_0^0=const$ which is a case of
given $\lambda $. The simplest algorithm for generating solutions is
presented. It is based on a master function and involves simple operations.
The same function is relevant also when $\nu $ is given.

In Sec.IV the case $\rho +p=0$ is solved completely. It represents a charged
generalization of de Sitter spacetime. One new solution is worked out as an
example. This case is generalized to include solutions with positive
pressure and an explicit example is given.

Sec.V deals with the case of given pressure which includes as a subcase
charged dust solutions (they have vanishing pressure). Efforts are
concentrated on the case $\left( \nu ,p\right) $ which is solved in general, 
$\lambda $ given as an explicit functional of the pressure. Conditions for
regular density are elucidated. The well-known case of regular CD is used as
an example to illustrate how the classification scheme works. Known
solutions are reviewed both in curvature and isotropic coordinates. A new
solution is given to demonstrate the features of a simple ansatz for $\nu $,
used with variations in the next sections, too. The most general CD solution
is derived. It is not regular at the centre but can form a 'halo' around a
typical solution with given pressure. An explicit example of such
triple-layered model with realistic properties is applied.

The case $\left( \nu ,q\right) $ is discussed in Sec.VI. It is the most
direct generalization of the uncharged case. An integral representation for $%
e^{-\lambda }$ is given. A series of ansatze for $\nu $ is introduced in
general, the first two cases are worked out explicitly. A new solution is
presented and some known results are corrected.

In Sec.VII the important case $\left( \nu ,n\right) $ of a fluid with linear
equation of state, involving the parameters $n$ and $p_0$, and given $\nu $
is investigated. Unlike the situation for uncharged fluid, the third main
equation is linear and of first order. It allows a generic integral
representation for $e^{-\lambda }$ which comprises also the cases $\left(
\nu ,q\right) $ and $\left( \nu ,\rho \right) $. The structure of the
density and its regularity are discussed which imply certain restrictions on 
$\nu $. The solutions are in some way electrified versions of the Einstein
static universe (ESU). A basic simplification technique leads to an infinite
series of models with realistic physical properties. Two of them are studied
in detail, including a model related to incoherent radiation. This kind of
fluid appears also as a special case of the integral representation. Another
special case is also soluble in elementary functions but has negative
pressure and is closer to the cases studied in Sec.IV. Finally, a third
special case, which is a twisted generalization of an obscure integrable
uncharged case, is shown to be ill-defined in the presence of charge.

Sec.VIII is dedicated to the case $\left( \nu ,\rho \right) $, closely
related to $\left( \nu ,n\right) $. A solution in terms of hypergeometric
functions is found when the density is constant. An algebraic ansatz for the
density, depending on two parameters, leads to realistic solution except for 
$q^2$ which turns out to be negative. The process of fine-tuning, however,
may give truly realistic solutions in the future.

In Sec.IX the three cases $\left( \rho ,q\right) $, $\left( \lambda ,\rho
\right) $ and $\left( \lambda ,q\right) $, which are closely related to each
other, are discussed. It represents a critical review of Refs. \cite
{six,twentytwo,twentythree,twentyfive,twentysix}. A unified scheme is
followed, which fully explores the connections between the different
subcases. Several new solutions are obtained which demonstrate how hard is
to escape the curse of negative pressure, which plagues purely
electromagnetic mass models. The solution of Wilson \cite{twentythree} is
formulated as a hypergeometric function.

In Sec.X the theme about a linear equation of state is raised once again,
this time in the frame of case $\left( \lambda ,n\right) $. First, a natural
generalization of the KT solution is obtained for constant $\lambda $, which
is quite different from the solution in Ref.\cite{forty}. Then the ansatz
for $\lambda $, which ensures that $T_0^0$ is constant, is studied in
detail. The general solution of the second order linear equation is
expressed once more through the hypergeometric function and degenerate cases
in elementary functions are sought for. Three different classes are studied.
One reduces to ESU. The second has nice orthogonal polynomial solutions for $%
e^{\nu /2}$ but with negative pressure, much like de Sitter solution which
is contained as a subcase. This unpleasant feature is also shared by a
trigonometric solution for the analog of incoherent radiation. The study of
the non-linear first order equation produces even more bizarre results.

Sec. XI contains some discussion and conclusions. In all sections old
solutions, known to the author, are reviewed and classified. New solutions
have been checked by computer for realistic properties.

\section{Main equations and classification}

The Einstein-Maxwell equations are written as 
\begin{equation}
\kappa T_0^0\equiv \kappa \rho +\frac{q^2}{r^4}=\frac{\lambda ^{\prime }}r%
e^{-\lambda }+\frac 1{r^2}\left( 1-e^{-\lambda }\right) ,  \label{two}
\end{equation}
\begin{equation}
\kappa p-\frac{q^2}{r^4}=\frac{\nu ^{\prime }}re^{-\lambda }-\frac 1{r^2}%
\left( 1-e^{-\lambda }\right) ,  \label{three}
\end{equation}
\begin{equation}
\kappa p+\frac{q^2}{r^4}=e^{-\lambda }\left( \frac{\nu ^{\prime \prime }}2-%
\frac{\lambda ^{\prime }\nu ^{\prime }}4+\frac{\nu ^{\prime 2}}4+\frac{\nu
^{\prime }-\lambda ^{\prime }}{2r}\right) ,  \label{four}
\end{equation}
where the prime means a derivative with respect to $r$ and $\kappa =8\pi
G/c^4$. We shall use units where $G=c=1$. The charge function is obtained by
integrating the charge density $\sigma $. We shall use, however, $q$ as a
primary object and then 
\begin{equation}
\kappa \sigma =\frac{2q^{\prime }}{r^2}e^{-\lambda /2}.  \label{five}
\end{equation}
Spherical symmetry allows only a radial electric field with potential $\phi $
given by 
\begin{equation}
F_{01}=\phi ^{\prime }=-\frac q{r^2}e^{\frac{\nu +\lambda }2}.  \label{six}
\end{equation}

Eq.(2) may be integrated by the introduction of the mass function 
\begin{equation}
M\left( r\right) =\frac \kappa 2\int_0^rr^2T_0^0dr=\frac \kappa 2%
\int_0^r\left( \rho +\frac{q^2}{\kappa r^4}\right) r^2dr,  \label{seven}
\end{equation}
and gives 
\begin{equation}
z\equiv e^{-\lambda }=1-\frac{2M}r.  \label{eight}
\end{equation}
This can be rewritten as 
\begin{equation}
\kappa \rho +\frac{q^2}{r^4}=\frac{2M^{\prime }}{r^2}=\frac 1{r^2}\left(
1-z-rz^{\prime }\right) ,  \label{nine}
\end{equation}
and constitutes the first of our main equations. The second is obtained as a
sum of Eqs. (2) and (3): 
\begin{equation}
\kappa \left( \rho +p\right) =\frac{e^{-\lambda }}r\left( \nu ^{\prime
}+\lambda ^{\prime }\right) =\frac zr\nu ^{\prime }-\frac{z^{\prime }}r.
\label{ten}
\end{equation}

The third main equation will not be Eq.(4) but another combination of
Eqs.(2)-(4) which varies from case to case. One can transform Eqs. (2)-(4)
into expressions for $p,q,\rho $%
\begin{equation}
2\kappa p=e^{-\lambda }\left( \frac{\nu ^{\prime \prime }}2-\frac{\lambda
^{\prime }\nu ^{\prime }}4+\frac{\nu ^{\prime 2}}4-\frac{\lambda ^{\prime }}{%
2r}+\frac{3\nu ^{\prime }}{2r}+\frac 1{r^2}\right) -\frac 1{r^2},
\label{eleven}
\end{equation}
\begin{equation}
\frac{2q^2}{r^4}=e^{-\lambda }\left( \frac{\nu ^{\prime \prime }}2-\frac{%
\lambda ^{\prime }\nu ^{\prime }}4+\frac{\nu ^{\prime 2}}4-\frac{\lambda
^{\prime }}{2r}-\frac{\nu ^{\prime }}{2r}-\frac 1{r^2}\right) +\frac 1{r^2},
\label{twelve}
\end{equation}
\begin{equation}
2\kappa \rho =e^{-\lambda }\left( -\frac{\nu ^{\prime \prime }}2+\frac{%
\lambda ^{\prime }\nu ^{\prime }}4-\frac{\nu ^{\prime 2}}4+\frac{5\lambda
^{\prime }}{4r}+\frac{\nu ^{\prime }}{2r}-\frac 1{r^2}\right) +\frac 1{r^2}.
\label{thirteen}
\end{equation}
These equations may be written as linear first-order equations for $z,$
suitable for the cases $\left( \nu ,q\right) ,\left( \nu ,p\right) $ and $%
\left( \nu ,\rho \right) $. Introducing $y=e^{\nu /2}$ we have 
\begin{equation}
\left( r^2y^{\prime }+ry\right) z^{\prime }=-2\left( r^2y^{\prime \prime
}-ry^{\prime }-y\right) z-2y+\frac{4q^2}{r^2}y,  \label{fourteen}
\end{equation}
\begin{equation}
\left( r^2y^{\prime }+5ry\right) z^{\prime }=-2\left( r^2y^{\prime \prime
}-ry^{\prime }+y\right) z+2y-4\kappa \rho r^2y,  \label{fifteen}
\end{equation}
\begin{equation}
\left( r^2y^{\prime }+ry\right) z^{\prime }=-2\left( r^2y^{\prime \prime
}+3ry^{\prime }+y\right) z+2y+4\kappa pr^2y.  \label{sixteen}
\end{equation}

In the uncharged case the prescription of an equation of state makes the
system of field equtions extremely difficult to solve. This is true even for
the simplest realistic linear equation of state 
\begin{equation}
p=n\rho -p_0,  \label{seventeen}
\end{equation}
where $n$ is a parameter taking values in the interval $[0,1]$ for
physically realistic solutions, while $p_0$ is a positive constant, allowing
the existence of a boundary of the fluid where $p=0$. When $p_0=0$ we obtain
the popular $\gamma $-law (with notation $n=\gamma -1$). In this case Eqs.
(2)-(4) with $q=0$ lead to an Abel differential equation of the second kind
independently from the approach or the coordinate system \cite{fortysix}. It
is soluble in few simple cases. Almost all of the known few solutions of the
more general Eq. (17) may be obtained by imposing a simple ansatz on $%
\lambda $ which makes the system overdetermined \cite{fortyseven}.
Therefore, it is surprising that in the more complex charged case fluids,
satisfying Eq. (17), are subjected to a linear equation, similar to Eqs.
(14)-(16). Plugging Eq. (17) into Eq. (10) yields 
\begin{equation}
\kappa \rho y=\frac 2{n+1}\left[ \frac{zy^{\prime }}r-2\left( \frac{%
z^{\prime }}r-\kappa p_0\right) y\right] .  \label{eighteen}
\end{equation}
Replacing this result in Eq. (15) leads to 
\begin{equation}
\left( r^2y^{\prime }+\frac{5n+1}{n+1}ry\right) z^{\prime }=-2\left(
r^2y^{\prime \prime }+\frac{3-n}{n+1}ry^{\prime }+y\right) z+2y-\frac{%
4\kappa p_0}{n+1}r^2y.  \label{nineteen}
\end{equation}
We call this case $\left( \nu ,n\right) $. It is a hybrid between $\left(
\nu ,\rho \right) $ and $\left( \nu ,p\right) $ because fixing the equation
of state, we fix the pressure in terms of the density. Eq. (17) is a
privileged one, due to its linearity. Another realistic equation of state,
namely the polytropic one, reads $p=n\rho ^{1+1/k}$. Replaced in Eq. (10) it
gives the relation 
\begin{equation}
\kappa \left( \rho +n\rho ^{1+1/k}\right) =\frac{2zy^{\prime }}{ry}-\frac{%
z^{\prime }}r.  \label{twenty}
\end{equation}
This causes the appearance of radicals in Eq. (15) at best, and to
non-integrable equations.

Eqs. (14)-(16) and (19) may be written in the general form 
\begin{equation}
gz^{\prime }=f_1z+f_0,  \label{twentyone}
\end{equation}
whose quadrature is 
\begin{equation}
z=e^F\left( C+H\right) ,  \label{twentytwo}
\end{equation}
\begin{equation}
F=\int \frac{f_1}gdr\text{,\qquad }H=\int e^{-F}\frac{f_0}gdr.
\label{twentythree}
\end{equation}
Here and in the following $C$ will denote a generic integration constant.
They may be written also as linear second-order differential equations for $%
y $, useful in the cases $\left( \lambda ,q\right) ,\left( \lambda ,\rho
\right) ,\left( \lambda ,p\right) $ and $\left( \lambda ,n\right) $: 
\begin{equation}
2r^2zy^{\prime \prime }+\left( r^2z^{\prime }-2rz\right) y^{\prime }+\left(
rz^{\prime }-2z+2-\frac{4q^2}{r^2}\right) y=0,  \label{twentyfour}
\end{equation}
\begin{equation}
2r^2zy^{\prime \prime }+\left( r^2z^{\prime }-2rz\right) y^{\prime }+\left(
5rz^{\prime }+2z-2+4\kappa \rho r^2\right) y=0,  \label{twentyfive}
\end{equation}
\begin{equation}
2r^2zy^{\prime \prime }+\left( r^2z^{\prime }+6rz\right) y^{\prime }+\left(
rz^{\prime }+2z-2-4\kappa pr^2\right) y=0,  \label{twentysix}
\end{equation}
\begin{equation}
2r^2zy^{\prime \prime }+\left( r^2z^{\prime }+2\frac{3-n}{n+1}rz\right)
y^{\prime }+\left( \frac{5n+1}{n+1}rz^{\prime }+2z-2+\frac{4\kappa p_0}{n+1}%
r^2\right) y=0.  \label{twentyseven}
\end{equation}
The coefficient before the second derivative is one and the same in all
cases. Eq. (24) is the generalization to the charged case \cite
{thirtyfive,thirtyseven} of the Wyman equation \cite{thirtyeight}.
Obviously, the case $n=-1$ is not covered by Eqs. (19), (27).

One can find first-order differential equations also for the cases $\left(
\lambda ,*\right) $, based on the well-known Tolman-Oppenheimer-Volkoff
(TOV) equation \cite{fortytwo,fortyeight}, generalized to the charged case 
\cite{twentyeight}. Let us first derive briefly TOV. Eq. (3) may be written
as an expression of $\nu ^{\prime }$ in terms of $p,q$ and $M$%
\begin{equation}
\nu ^{\prime }=\frac{2M+\kappa pr^3-q^2/r}{r\left( r-2M\right) }.
\label{twentyeight}
\end{equation}
Now, let us take the following combination of equations $\left( 3\right)
^{\prime }+\frac 2r\left( 3\right) -\frac 2r\left( 4\right) $. After some
cancellations it gives 
\begin{equation}
\kappa p^{\prime }=-\kappa \left( \rho +p\right) \frac{\nu ^{\prime }}2+%
\frac{\left( q^2\right) ^{\prime }}{r^4}.  \label{twentynine}
\end{equation}
Inserting Eq. (28) into Eq. (29) yields the generalized TOV equation. We can
trade $q$ in Eqs.(28)-(29) for $\rho $ and $\lambda $ by using Eq. (9). The
result is a Riccati equation for $Y\equiv \kappa \left( \rho +p\right) $%
\begin{equation}
Y^{\prime }=-\frac r2e^\lambda Y^2+\frac{\lambda ^{\prime }}2Y-4\kappa \frac 
\rho r+\frac 2{r^4}\left( r^2M^{\prime }\right) ^{\prime }.  \label{thirty}
\end{equation}
It marks another way to solve the cases $\left( \lambda ,\rho \right)
,\left( \lambda ,n=-1\right) $. Its solution yields for the pressure $\kappa
p=Y-\kappa \rho $. Then $\nu $ is found from Eq. (10), while $q$ is found
from Eq. (9). Eq. (30) is a non-linear, but first-order companion of Eq.
(15). Using Eq. (9) we find again a Riccati equation for the case $\left(
\lambda ,q\right) $%
\begin{equation}
Y^{\prime }=-\frac r2e^\lambda Y^2+\frac{\lambda ^{\prime }}2Y+\frac{4q^2}{%
r^5}-\frac{8M^{\prime }}{r^3}+\frac 2{r^4}\left( r^2M^{\prime }\right)
^{\prime }.  \label{thirtyone}
\end{equation}
From the definition of $Y$ equations for the cases $\left( \lambda ,p\right) 
$ and $\left( \lambda ,n\right) $ follow 
\begin{equation}
Y^{\prime }=-\frac r2e^\lambda Y^2+\left( \frac{\lambda ^{\prime }}2-\frac 4r%
\right) Y+4\kappa \frac pr+\frac 2{r^4}\left( r^2M^{\prime }\right) ^{\prime
},  \label{thirtytwo}
\end{equation}
\begin{equation}
Y^{\prime }=-\frac r2e^\lambda Y^2+\left( \frac{\lambda ^{\prime }}2-\frac 4{%
\left( n+1\right) r}\right) Y-\frac{4\kappa p_0}{\left( n+1\right) r}+\frac 2%
{r^4}\left( r^2M^{\prime }\right) ^{\prime }.  \label{thirtythree}
\end{equation}
In all cases Riccati equations are obtained with the same coefficients
before $Y^{\prime }$ and $Y^2$. They may be written as 
\begin{equation}
Y^{\prime }=-\frac r2e^\lambda Y^2+f_2Y+f_3,  \label{thirtyfour}
\end{equation}
where $f_2$ and $f_3$ are functions of $r$, differing from case to case. An
arbitrary Riccati equation may be transformed into a linear second-order
equation. The linearization of Eq. (34) is done by the change of variables $%
u=e^{\lambda /2}y$ and reads 
\begin{equation}
u^{\prime \prime }-\left( \frac 1r+\lambda ^{\prime }+f_2\right) u^{\prime }-%
\frac r2e^\lambda f_3u=0.  \label{thirtyfive}
\end{equation}
At this point $f_3$ contains $\lambda ^{\prime \prime }$ which does not
cancel. When a passage from $u$ to $y$ is done, $\lambda ^{\prime \prime }$
cancels and we obtain exactly Eqs. (24)-(27). In this process we have
exchanged $Y$, which is the sum of the pressure and the density, for $y$
which is part of the metric.

So far we have reformulated the original system of Eqs. (2)-(4) into Eqs.
(9), (10) and a third equation, presented in many different forms, adapted
to the cases of the proposed classification. We have briefly discussed the
cases $\left( \lambda ,\nu \right) ,\left( \lambda ,*\right) ,\left( \nu
,*\right) $. The three remaining cases $\left( \rho ,p\right) ,\left( \rho
,q\right) $ and $\left( p,q\right) $ are the most natural ones since one
prescribes two of the fluid characteristics, hoping that the third one and
the metric will be regular and reasonable. The case $\left( \rho ,q\right) $
is easily reduced to $\left( \lambda ,q\right) $ because of Eq. (9). In the
case $\left( \rho ,p\right) $, $Y$ is also known and Eq. (30) becomes an
equation for $M$%
\begin{equation}
\frac 2{r^4}\left( r^2M^{\prime }\right) ^{\prime }+\frac{rM^{\prime }-M}{%
r\left( r-2M\right) }Y-\frac{r^2Y^2}{2\left( r-2M\right) }-Y^{\prime }-\frac{%
4\kappa \rho }r=0.  \label{thirtysix}
\end{equation}
This is an intricate, non-linear, second-order equation, which is not
simpler than the TOV equation. It seems that it can be dealt with only
numerically.

The third main equation for the case $\left( p,q\right) $ is obtained in the
following way. Let us replace the density in the TOV equation (29) with its
expression from Eq. (9). After some tedious manipulations, the following
equation for $M$ is found 
\begin{equation}
\left( M+g_0\right) M^{\prime }=f_4M+f_5,  \label{thirtyseven}
\end{equation}
\begin{equation}
g_0=\frac \kappa 2pr^3-\frac{q^2}{2r},  \label{thirtyeight}
\end{equation}
\begin{equation}
f_4=\kappa r^3p^{\prime }-\frac{\left( q^2\right) ^{\prime }}r-\frac \kappa 2%
r^2p+\frac{q^2}{2r^2},  \label{thirtynine}
\end{equation}
\begin{equation}
2f_5=-r^4\left( \kappa p^{\prime }-\frac{\left( q^2\right) ^{\prime }}{r^4}%
\right) -r^2\left( \kappa p-\frac{q^2}{r^4}\right) \left( \frac \kappa 2pr^3-%
\frac{q^2}{2r}\right) .  \label{forty}
\end{equation}
There is a standard procedure for the solution of such equations \cite
{fortysix,fortynine}. It consists of two changes of variables which bring
them to the Abel equation of the second kind 
\begin{equation}
\omega \omega _\zeta -\omega =f\left( \zeta \right) ,  \label{fortyone}
\end{equation}
where $\omega =M+g_0$ while 
\begin{equation}
\zeta =g_0+\int f_4dr,  \label{fortytwo}
\end{equation}
\begin{equation}
f\left( \zeta \right) =\frac{f_5-g_0f_4}{g_0^{\prime }+f_4}.
\label{fortythree}
\end{equation}
Its integrable cases are few, depend on the shape of $f\left( \zeta \right) $
and are tabulated in Ref. \cite{fortynine}.

As a whole, the most attractive are the mixed cases $\left( \nu ,*\right) $
where one fluid characteristic and one metric component are specified. They
lead to the most simple Eqs. (14)-(16) and (19).

The five functions which describe the fluid together with its gravitational
field should satisfy some physical requirements. Eqs.(7)-(8) show that at
the centre $M\left( 0\right) =0$ and $e^\lambda =1$. The density and
pressure should be positive and monotonously decreasing towards the
boundary. It is obvious that $q^2$ should be positive, too. The boundary $%
r_0 $ of the fluid sphere is determined by the relation $p\left( r_0\right)
=0$ where a junction to the Reissner-Nordstr\"om (RN) metric 
\begin{equation}
e^\nu =e^{-\lambda }=1-\frac{2m}r+\frac{e^2}{r^2}  \label{fortyfour}
\end{equation}
should be performed. The metric and $\nu ^{\prime }$ must be continuous
there. This leads to the expressions 
\begin{equation}
\frac m{r_0}=1-e^\nu \left( 1+\frac{r_0\nu ^{\prime }}2\right) =\frac{%
M\left( r_0\right) }{r_0}+\frac{q^2\left( r_0\right) }{2r_0^2},
\label{fortyfive}
\end{equation}
\begin{equation}
\frac{e^2}{r_0^2}=1-e^\nu \left( 1+r_0\nu ^{\prime }\right) =\frac{q^2\left(
r_0\right) }{r_0^2}.  \label{fortysix}
\end{equation}
The condition $e=q\left( r_0\right) $ follows from the vanishing of the
pressure and vice versa. In fact, Eq. (10) gives at $r_0$%
\begin{equation}
z^{\prime }=z\nu ^{\prime }-\kappa r_0\rho \left( r_0\right) .
\label{fortyseven}
\end{equation}
Replacing this in Eq. (9) and taking into account that $e^\nu \left(
r_0\right) =e^{-\lambda }\left( r_0\right) $, we get a proof of the
assertion.

Solutions, satisfying the above conditions are called in the following
physically realistic. This is done in agreement with the majority of the
reviewed papers, and their satisfaction is already rather non-trivial. Some
other, more stringent requirements, are discussed in Refs. \cite
{fifty,fiftyone,fiftytwo}. The most important of them is $0\leq \frac{dp}{%
d\rho }\leq 1$, which means that the speed of sound is positive and causal,
i.e., not greater than the speed of light. One has control over this
characteristic in CD and models with linear equation of state. The case $%
\left( \nu ,q\right) $, which is a generalization of uncharged solutions,
possesses their behavior for small $q$. In other cases the expressions for
density and pressure may be so complicated that the fulfilment of this
requirement is hard to estimate. We do not discuss these additional
conditions in order to contain the present paper in a reasonable volume and
prevent its dissociation into a set of separate publications. In many cases
the examples have free parameters, or such can be added easily, so
additional fine-tuning may be done.

\section{The cases $\left( \lambda ,\nu \right) $ and constant $T_0^0$}

In the case $\left( \lambda ,\nu \right) $ Eqs. (11)-(13) should be used to
find $q,p$ and $\rho $. This is the simplest case but control over pressure
and density is completely lost and one must proceed by trial and error.
Krori and Barua \cite{thirtythree} have given the solution 
\begin{equation}
\lambda =a_1r^2,  \label{fortyeight}
\end{equation}
\begin{equation}
\nu =a_2r^2+a_3,  \label{fortynine}
\end{equation}
where $a_i$ denote generic constants of a known solution. They are fixed by
the junction conditions. The solution is non-singular and the positivity
conditions are satisfied.

Eqs.(7)-(8) show that the generalized Schwarzschild condition $T_0^0=const$
determines $\lambda $%
\begin{equation}
e^{-\lambda }=1-ar^2,  \label{fifty}
\end{equation}
where $a$ is a positive constant. We have the freedom to choose one more
function. In Ref. \cite{six} the condition $p=0$ was further imposed. We
shall review this solution in Sec.V. In Ref. \cite{twentyone} some relations
between $\phi $ and $\nu $ were utilized. They can be arbitrary in the
perfect fluid case, so we do not base the classification on such criteria.
One should use instead Eqs. (24)-(27) or Eqs. (30)-(33). The simplest
expressions are obtained in the case $\left( \lambda ,Y\right) $ when Eq.
(30) is used to determine $\rho $%
\begin{equation}
4\kappa \rho =-rY^{\prime }-\frac{r^2Y^2}{2\left( 1-ar^2\right) }+\frac{arY}{%
1-ar^2}+12a.  \label{fiftyone}
\end{equation}
Then $\kappa p=Y-\kappa \rho $ and 
\begin{equation}
q^2=r^4\left( 3a-\kappa \rho \right) =\frac{r^5}4\left[ Y^{\prime }+\frac{%
rY\left( Y-2a\right) }{2\left( 1-ar^2\right) }\right] .  \label{fiftytwo}
\end{equation}
The function $\nu $ is determined by a simple integration from Eq. (10) 
\begin{equation}
\nu ^{\prime }=\frac r{1-ar^2}\left( Y-2a\right) .  \label{fiftythree}
\end{equation}
This is an advantage over the second-order equations (24)-(27). Eqs.
(50)-(53) solve the problem in a minimal algebraic way. Several positivity
conditions follow for the master function $Y$ and $\rho $: $3a>\kappa \rho
,Y>2a,Y>\kappa \rho ,Y^{\prime }<0$.

In order to illustrate how the above scheme works one may take any of the
solutions elaborated in Ref. \cite{twentyone}, extract $Y$ and check how the
above equations and inequalities are satisfied. The simplest case has 
\begin{equation}
Y=2a+a_1\left( 1-ar^2\right) ^{1/2}.  \label{fiftyfour}
\end{equation}

Of course, in any of the cases $\left( \lambda ,*\right) $ we can study the
subcase given by Eq. (50). This will be done in the following sections.

A similar problem arises when $\nu $ is given. Which function should be
prescribed in addition in order to have the simplest algorithm for
generating solutions? Again, $Y$ is the best choice. This is seen by taking
the difference of Eqs. (25),(26) or reformulating Eq. (10) 
\begin{equation}
yz^{\prime }-2y^{\prime }z+ryY=0.  \label{fiftyfive}
\end{equation}
This equation is much simpler than any of Eqs. (14)-(16).

\section{Charged de Sitter solutions and their generalization}

In the uncharged case this is a special integrable case with linear equation
of state (17) and $n=-1,$ $p_0=0$ which is equivalent to the de Sitter
solution. The charged case is also completely solvable and falls in the case 
$\left( \lambda ,Y=0\right) $ according to our classification. The third
main equation is Eq. (30) which becomes 
\begin{equation}
\kappa \rho =\frac{\left( r^2M^{\prime }\right) ^{\prime }}{2r^3}=\frac 1{%
2r^2}\left( 2M^{\prime }+rM^{\prime \prime }\right) .  \label{fiftysix}
\end{equation}
Eq. (10) gives $\nu =-\lambda $ which is a feature also of the exterior RN
solution. Eq. (9) yields 
\begin{equation}
q^2=-\frac{r^5}2\left( \frac{M^{\prime }}{r^2}\right) ^{\prime }=\frac r2%
\left( 2M^{\prime }-rM^{\prime \prime }\right) .  \label{fiftyseven}
\end{equation}
Thus, when $M$ is given, all other unknowns follow from simple formulae. Eq.
(57) allows also one to take $q$ as a basis: 
\begin{equation}
\kappa \rho =2a_0-\frac{q^2}{r^4}-4\int_0^r\frac{q^2}{r^5}dr,
\label{fiftyeight}
\end{equation}
\begin{equation}
M=-2\int_0^r\bar r^2\int_0^{\bar r}\frac{q^2}{\breve r^5}d\breve rd\bar r+%
\frac{a_0}3r^3.  \label{fiftynine}
\end{equation}
Here $a_0$ is some positive constant and clearly $q=r^{2+\varepsilon
}q_0\left( r\right) $ where $\varepsilon >0$ and $q_0\left( 0\right) =const$%
. Eqs. (58)-(59) demonstrate the process of 'electrification' of de Sitter
space. The bigger the charge function, the lower the density until some
point $r_0$ is reached where $\rho \left( r_0\right) =0$ and consequently
the fluid sphere acquires a boundary. We already know that $M>0$. Eqs.
(56)-(57) show that $\rho $ and $q^2$ are both positive only when $M^{\prime
}>0$ and $2M^{\prime }\geq r\left| M^{\prime \prime }\right| $. The equality
holds at the boundary, where $M^{\prime \prime }\left( r_0\right) <0$. The
charge and the mass of the solution follow from the junction conditions
(45),(46) 
\begin{equation}
e^2=2r_0^2M^{\prime }\left( r_0\right) ,  \label{sixty}
\end{equation}
\begin{equation}
m=M\left( r_0\right) +r_0M^{\prime }\left( r_0\right) .  \label{sixtyone}
\end{equation}

Obviously, there is an abundance of solutions since $M$ and its first two
derivatives have to satisfy few simple inequalities. Three solutions are
known in the literature. The $T_0^0=const$ condition leads to $q=0$ in the
interior and consequently to the de Sitter solution, which has constant
density. One can introduce, however, a surface charge $\sigma \sim \delta
\left( r-r_0\right) $ which gives $q=\theta \left( r-r_0\right) e$, $\rho
=\rho _0$ and 
\begin{equation}
z=1-\frac \kappa 3\rho _0r^2.  \label{sixtytwo}
\end{equation}
This is the solution of Cohen and Cohen \cite{twentyseven,twentynine}.
Another solution \cite{twentyeight} has 
\begin{equation}
M=\frac{\kappa ^2}{360}\sigma _0^2r^3\left( 5a_1^2-2r^2\right) ,
\label{sixtythree}
\end{equation}
\begin{equation}
q=\frac \kappa 6\sigma _0r^3,  \label{sixtyfour}
\end{equation}
\begin{equation}
\rho =\frac \kappa {12}\sigma _0^2\left( a_1^2-r^2\right) ,
\label{sixtyfive}
\end{equation}
\begin{equation}
z=1-\frac{\kappa ^2}{36}\sigma _0^2a_1^2r^2+\frac{\kappa ^2}{90}\sigma
_0^2r^4.  \label{sixtysix}
\end{equation}
Here $\sigma _0$ enters the charge density $\sigma =\sigma _0e^{-\lambda /2}$%
, while $a_1$ is related to $a_0=\frac{\kappa ^2}{24}\sigma _0^2a_1$. This
is one of the electromagnetic models of the electron. When $\sigma
_0\rightarrow 0$ we have $\rho ,p,q,\lambda ,\nu \rightarrow 0$ due to $%
a_0\rightarrow 0$. The limit is flat spacetime. We have shown, however, that 
$a_0$ is not obliged to vanish when $q=0$ so that, in general, the case $%
\left( \lambda ,Y=0\right) $ is an electric generalization of de Sitter
spacetime. The natural generalization of flat spacetime is the CD solution,
which will be discussed in the next section.

A third solution has been found by Gautreau \cite{thirty}. It has 
\begin{equation}
2M=c_1^2r-\frac{c_1^2a_2^2}r\sin ^2\frac r{a_2},  \label{sixtyseven}
\end{equation}
\begin{equation}
z=1-c_1^2+\left( \frac{c_1a_2}r\sin \frac r{a_2}\right) ^2,
\label{sixtyeight}
\end{equation}
\begin{equation}
\kappa \rho =\frac{c_1^2}{r^2}\sin ^2\frac r{a_2},  \label{sixtynine}
\end{equation}
\begin{equation}
\phi =c_1+\frac{c_1a_2}r\sin \frac r{a_2},  \label{seventy}
\end{equation}
and $q=-r^2\phi ^{\prime }$. Many other solutions are possible and we give
one simple example, a variation of the solution in Ref.\cite{twentyeight}: 
\begin{equation}
M=a\left( r^3-r^4\right) ,  \label{seventyone}
\end{equation}
\begin{equation}
\kappa \rho =2a\left( 3-5r\right) ,  \label{seventytwo}
\end{equation}
\begin{equation}
q^2=2ar^5,  \label{seventythree}
\end{equation}
\begin{equation}
z=1-2ar^2+2ar^3.  \label{seventyfour}
\end{equation}
The junction at $r_0=3/5$ gives $m=ar_0^3$, $e^2=2ar_0^5$. $M$ satisfies all
necessary inequalities and the density is positive and decreasing.

The case $Y=0$ is easy because Eq. (30) collapses into the simple relation
(56). A generalization can be made by taking Eq.(56) as a basis. This
constitutes the special case $\left( \lambda ,\rho \left( \lambda \right)
\right) $. Then $q$ is given again by Eq. (57), while Eq. (30) becomes 
\begin{equation}
Y^{\prime }=-\frac r2e^\lambda Y^2+\frac{\lambda ^{\prime }}2Y.
\label{seventyfive}
\end{equation}
This is a Bernoulli equation and, unlike the Riccati equation, it is readily
soluble in quadratures. Its general solution gives an expression for the
pressure 
\begin{equation}
\kappa p=\frac{e^{\lambda /2}}{C+\frac 12\int e^{3\lambda /2}rdr}-\kappa
\rho ,  \label{seventysix}
\end{equation}
where $C^{-1}=Y\left( 0\right) $. When $C\rightarrow \infty $ we return to
the previous case $Y=0$, the trivial solution of Eq. (75). Fortunately, Eq.
(10) can be integrated explicitly too and a closed expression is obtained
for $\nu $%
\begin{equation}
e^{\nu /2}=A^{-1}e^{-\lambda /2}\left( 1+\frac 1{2C}\int_0^re^{3\lambda
/2}rdr\right) ,  \label{seventyseven}
\end{equation}
\begin{equation}
A=1+\frac 1{2C}\int_0^{r_0}e^{3\lambda /2}rdr.  \label{seventyeight}
\end{equation}
The second equation follows from the junction conditions. Eqs.
(56),(57),(76)-(78) provide the generalization of the case $Y=0$. Thus,
every function $M$ leads to two solutions: one with trivial $Y$ and one with
non-trivial $Y$, satisfying Eq. (75). The trivial solution has the
disadvantage of negative pressure. Eq.(76) suggests that in the non-trivial
case solutions with positive pressure may exist.

Let us take Eq. (63) with somewhat different constants 
\begin{equation}
M=br^3\left( 1-br^2\right) .  \label{seventynine}
\end{equation}
Then 
\begin{equation}
\kappa \rho =3b\left( 2-5bx\right) ,  \label{eighty}
\end{equation}
\begin{equation}
q^2=5b^2x^3,  \label{eightyone}
\end{equation}
\begin{equation}
\kappa p=\frac 1{C\left( 1-2bx+2bx^2\right) ^{1/2}-\frac 1{4b}+\frac x2}%
+15b^2x-6b,  \label{eightytwo}
\end{equation}
where $x=r^2$ and $b>0$. When $x<2/5b$ the density is positive and
decreasing. The pressure is positive at the centre when $1/4b<C<5/12b$. Let
us choose $b=0.01$ and $C=40$. Then $p$ has a maximum at $x_0=0.3$ and a
root at $x_0=2.5$. This is a semi-realistic interior solution with $%
e^2=5b^2x_0^3$ and mass given by Eq. (61).

\section{The case of given pressure. Charged dust}

In this section we discuss the cases $\left( \nu ,p\right) $ and $\left(
\lambda ,p\right) $. The pressure is considered a known positive function
which decreases monotonously outwards and vanishes at the boundary of the
fluid sphere. The simplest case is $p=0$ (the only reasonable case with
constant pressure, unlike the cases with constant density). This represents
charged dust. The case $\left( \nu ,p\right) $ is much easier since the
third main equation (16) is linear and first-order. Something more, in Eq.
(21) $f_1=-2g^{\prime }$. This allows to obtain a compact expression for $z$%
\begin{equation}
z=\frac 1{\left( 1+\frac{r\nu ^{\prime }}2\right) ^2}+\frac C{r^2e^\nu
\left( 1+\frac{r\nu ^{\prime }}2\right) ^2}+\frac{4\kappa }{r^2e^\nu \left(
1+\frac{r\nu ^{\prime }}2\right) ^2}\int_0^r\left( 1+\frac{r\nu ^{\prime }}2%
\right) e^\nu r^3pdr.  \label{eightythree}
\end{equation}
The knowledge of $\nu $ allows to satisfy two of the junction conditions by
choosing a function continuous at $r_0$ together with its derivative.
Eqs.(9)-(10) provide an expression for $q$ in terms of the known $\nu ,p$
and with $z\left( \nu ,p\right) $ from Eq. (83) 
\begin{equation}
\frac{q^2}{r^2}=1-z\left( 1+r\nu ^{\prime }\right) +\kappa pr^2.
\label{eightyfour}
\end{equation}
The density follows from Eq. (10) 
\begin{equation}
\kappa \rho =\frac zr\left( \nu ^{\prime }-\frac{z^{\prime }}z\right)
-\kappa p.  \label{eightyfive}
\end{equation}
Its structure is clarified when $z^{\prime }$ is replaced by $\nu ^{\prime }$%
, using Eq. (83) 
\begin{equation}
\kappa \rho =\frac{2z}{r^2}-\frac 2{r^2\left( 1+\frac{r\nu ^{\prime }}2%
\right) }+\frac{2z\nu ^{\prime }}r+\frac{z\nu ^{\prime }}{r\left( 1+\frac{%
r\nu ^{\prime }}2\right) }+\frac{z\nu ^{\prime \prime }}{1+\frac{r\nu
^{\prime }}2}-\frac{4\kappa p}{1+\frac{r\nu ^{\prime }}2}-\kappa p.
\label{eightysix}
\end{equation}
When $r\rightarrow 0$, $z\rightarrow 1$ and the third term on the right
produces a pole unless $\nu ^{\prime }\rightarrow 2\nu _0r$, which means $%
\nu ^{\prime \prime }\rightarrow 2\nu _0$. Then $r\nu ^{\prime }\rightarrow
0 $ and the poles in the first two terms cancel. The last two terms approach
negative constants. In order to compensate them and have a positive density
at the centre, $\nu _0$ must be a positive constant and the inequality $8\nu
_0>5\kappa p\left( 0\right) $ should hold. Another consequence is that $%
e^\nu $ is an increasing function in the vicinity of the centre. Since $%
e^{-\lambda }$ is a decreasing function which starts from $1$ and equals $%
e^\nu $ at $r_0$, we have $e^\nu \left( 0\right) <1$. Eq. (83) shows that $z$
has a pole at $r=0$ unless $C=0$.

Now we can understand the physical meaning of the terms in $z$. The third
term represents the contribution from the pressure to the metric. When $p=0$%
, $z$ is still non-trivial and represents the general CD solution. It has a
pole at the beginning of the coordinates and should not be used there. When $%
C=0$ only the first term remains and this is the regular CD solution in
curvature coordinates \cite{one}. The intricate proofs that this is the most
general regular solution \cite{two,four} are replaced here by the obvious
condition $C=0$. The second term in Eq. (83) may be induced in principle by
the third if the lower limit of the integral is changed. The first term may
be absorbed by the third if one uses the identity 
\begin{equation}
\left( 1+\frac{r\nu ^{\prime }}2\right) e^\nu r=\frac 12\left( r^2e^\nu
\right) ^{\prime },  \label{eightyseven}
\end{equation}
and makes the shift $\kappa p\rightarrow \kappa p+\frac 1{2r^2}$. The
expression under the integral in Eq. (83) is bell-shaped, since it vanishes
both at zero and $r_0$ and is positive in-between.

Let us discuss in detail first the case of regular CD. It provides an
excellent illustration of the classification scheme, adopted in the present
paper. In the case $\left( \nu ,p=0\right) $ Eqs. (5)-(6) and (83)-(86) give 
\begin{equation}
e^\lambda =\left( 1+\frac{r\nu ^{\prime }}2\right) ^2,  \label{eightyeight}
\end{equation}
\begin{equation}
q=\frac{r^2\nu ^{\prime }}{2\left( 1+\frac{r\nu ^{\prime }}2\right) },
\label{eightynine}
\end{equation}
\begin{equation}
\kappa \rho =\kappa \sigma =\frac{r\nu ^{\prime \prime }+2\nu ^{\prime }+%
\frac 12r\nu ^{\prime 2}}{r\left( 1+\frac{r\nu ^{\prime }}2\right) ^3},
\label{ninety}
\end{equation}
\begin{equation}
\phi =-e^{\nu /2}+\phi _0.  \label{ninetyone}
\end{equation}
For the case $\left( \lambda ,p=0\right) $ we should use Eq. (88) to express
everything via $\lambda $%
\begin{equation}
\nu ^{\prime }=\frac 2r\left( e^{\lambda /2}-1\right) ,  \label{ninetytwo}
\end{equation}
\begin{equation}
q=r\left( 1-e^{-\lambda /2}\right) ,  \label{ninetythree}
\end{equation}
\begin{equation}
\kappa \rho =\kappa \sigma =\frac 2{r^2}\left( e^{-\lambda /2}-e^{-\lambda }+%
\frac r2\lambda ^{\prime }e^{-\lambda }\right) .  \label{ninetyfour}
\end{equation}
These formulae involve only the first derivative of $\lambda $. The case $%
\left( q,p=0\right) $ is explicitly solvable, unlike the general case $%
\left( q,p\right) $: 
\begin{equation}
e^\lambda =\frac{r^2}{\left( r-q\right) ^2},  \label{ninetyfive}
\end{equation}
\begin{equation}
M=q-\frac{q^2}{2r},  \label{ninetysix}
\end{equation}
\begin{equation}
\nu ^{\prime }=\frac{2q}{r\left( r-q\right) },  \label{ninetyseven}
\end{equation}
\begin{equation}
\kappa \rho =\kappa \sigma =\frac 2{r^3}\left( r-q\right) q^{\prime }.
\label{ninetyeight}
\end{equation}
The condition $q<r$ must hold. These formulae clearly demonstrate the
electrification of flat-space, which is the trivial dust solution in the
uncharged case. If we put $p=0$ in Eqs. (37)-(40) we still obtain a
complicated Abel equation. One can check that $M$, given by Eq. (96),
satisfies it. The most complicated is the $\left( \rho ,p=0\right) $ case.
If we introduce $w$ via 
\begin{equation}
e^\lambda =w^{-2},  \label{ninetynine}
\end{equation}
then Eqs. (92),(93) give 
\begin{equation}
\nu ^{\prime }=\frac{2\left( 1-w\right) }{rw},  \label{h}
\end{equation}
\begin{equation}
q=r\left( 1-w\right) ,  \label{hone}
\end{equation}
while Eq. (94) becomes an equation for $w$%
\begin{equation}
rww^{\prime }=-w^2+w-\frac \kappa 2r^2\rho .  \label{htwo}
\end{equation}
This is an Abel differential equation of the second kind like Eq. (36). The
procedure for its solution was described briefly after Eq. (40) and brings
it to the canonical form \cite{fortysix,fortynine} 
\begin{equation}
WW^{\prime }-W=-\frac \kappa 2r^3\rho ,  \label{hthree}
\end{equation}
where $W=wr$. The set of density profiles, leading to integrable $W$ is very
restricted.

Several explicit dust solutions in curvature coordinates have been given.
Efinger \cite{five} studied the simple case $e^{-\lambda }=4/9$ in the
presence of a cosmological constant. When it is zero we have $q=r/3$ and
singular $e^\nu =a_1r$, $\rho =\sigma =4/9\kappa r^2$, $F_{10}=\frac 12\sqrt{%
a_1}r^{-1/2}$. Florides illustrated his general discussion \cite{four} with
a power-law solution $q=a_2r^{a_3+3}$, 
\begin{equation}
e^\lambda =\left( 1-a_2r^{a_3+2}\right) ^{-2},  \label{hfour}
\end{equation}
\begin{equation}
e^\nu =\left( 1-a_2r^{a_3+2}\right) ^{-\frac 2{a_3+2}},  \label{hfive}
\end{equation}
\begin{equation}
\rho =\sigma =\sigma _0\left( \frac r{b_1}\right) ^{a_3}e^{-\lambda /2},
\label{hsix}
\end{equation}
where $a_2=\frac{\kappa \sigma _0}{2\left( a_3+3\right) b_1^{a_3}}$.
Finally, in Ref.\cite{six} the case of constant $T_0^0$ was discussed. It
leads to Eq. (50) and Eqs. (92)-(94) give 
\begin{equation}
\nu =2\ln \left[ a_4\frac{1-\left( 1-a^2r^2\right) ^{1/2}}{a^2r^2}\right] ,
\label{hseven}
\end{equation}
\begin{equation}
q^2=r^2\left[ 2-a^2r^2-2\left( 1-a^2r^2\right) ^{1/2}\right] ,
\label{height}
\end{equation}
\begin{equation}
\kappa \rho =4a^2-\frac 2{r^2}\left[ 1-\left( 1-a^2r^2\right) ^{1/2}\right] .
\label{hnine}
\end{equation}
We use the known solutions just to illustrate the general ideas and formulae
and do not elaborate on their physical significance or details of the
junction conditions, which may be found in the original references.

As mentioned in the introduction, most work on CD has been done in isotropic
coordinates 
\begin{equation}
ds^2=U^{-2}dt^2-U^2\left( dr^2+r^2d\Omega ^2\right) ,  \label{hten}
\end{equation}
where the simple relation $\phi =\pm U^{-1}$ holds and the only equation to
be solved is 
\begin{equation}
U^{\prime \prime }+\frac 2rU^{\prime }=-\frac \kappa 2U^3\rho .
\label{heleven}
\end{equation}
When $U$ is given, we have the mixed case $\left( U,p=0\right) $ and the
density is readily determined from Eq. (111). The function $%
U^{-2}=a_1r^2+a_2 $ was used in Refs. \cite{ten,fifteen}. Another, more
general function 
\begin{equation}
U=1+\frac{a_3}{r_0}+\frac{a_3\left( r_0^k-r^k\right) }{kr_0^{k+1}}
\label{htwelve}
\end{equation}
was studied for $k=2$ in Ref. \cite{twelve}, for $k=4$ in Ref. \cite{eleven}
and for general $k$ in Ref. \cite{thirteen}. Recently, the function $U=a_4%
\frac{\sin a_5r}r$ was studied \cite{sixteen,seventeen}.

When $\rho $ is given, it is convenient to transform Eq. (111) into 
\begin{equation}
U_{XX}=-\frac{\kappa \rho }2\frac{U^3}{X^4},  \label{hthirteen}
\end{equation}
where $X=1/r$. We may consider $\rho \left( U\right) $ as a known function,
chosen to simplify this equation. One possibility, leading to Bessel
functions, is $\rho =a_6/U^2$ \cite{sixteen,seventeen}. Other choices lead
to two-dimensional integrable models, like the sine-Gordon equation \cite
{eighteen}. The metric (110) describes also the regular static CD. Then the
case of constant fluid density $\rho _0$ is soluble in $cn$, one of the
Jacobi elliptic functions, but the solution cannot be spherically symmetric.
In the spherically symmetric case Eq. (113) becomes an Emden-Fowler
equation, whose integrable cases are tabulated in Ref. \cite{fortynine}.
When the power of $U$ is $3$ like in Eq. (113), the integrable cases have $%
X^3$ or $X^6$ and not $X^4$. We can make the transformation $\psi =-U/r$, $%
Z=U_r-U/r$ and obtain 
\begin{equation}
ZZ_\psi -Z=-\frac{\kappa \rho _0}2\psi ^3.  \label{hfourteen}
\end{equation}
This equation coincides exactly with the Abel equation (103) in curvature
coordinates with constant density and is non-integrable too.

Let us go back to curvature coordinates (1) which are more convenient when
one wants to study also the case of non-vanishing pressure. From the
considerations about the regularity of $\rho $, given after Eq. (86), it
follows that the simplest function $\nu $ would be $\nu =\nu _0r^2$. This
choice coincides with Eq. (49) and leads to exponential functions and
probably to the error function in Eq. (83). We shall choose a different
option 
\begin{equation}
e^\nu =a+br^2,  \label{hfifteen}
\end{equation}
where $0<a<1$ and $b>0$. This ansatz has been thoroughly studied in the
uncharged case \cite{fifty,fiftytwo}. The CD solution build on it has 
\begin{equation}
z=\left( \frac{a+br^2}{a+2br^2}\right) ^2,  \label{hsixteen}
\end{equation}
\begin{equation}
q=\frac{br^3}{a+2br^2},  \label{hseventeen}
\end{equation}
\begin{equation}
\kappa \rho =\frac{2b\left( a+br^2\right) \left( 3a+2br^2\right) }{\left(
a+2br^2\right) ^3}.  \label{heighteen}
\end{equation}
Density is monotonously decreasing. The junction conditions (45)-(46) give 
\begin{equation}
e=\frac{br_0^3}{a+2br_0^2},  \label{hnineteen}
\end{equation}
\begin{equation}
m=r_0\left( 1-a-2br_0^2\right) .  \label{htwenty}
\end{equation}
The condition $z=e^\nu $ at $r_0$ supplies the relation 
\begin{equation}
r_0^2=\frac{1-4a+\sqrt{1+8a}}{8b},  \label{htwentyone}
\end{equation}
which ensures the obligatory $e=m$. The r.h.s. is positive when $a<1$, which
is the case.

The main disadvantage of regular CD solutions is that they possess a fixed
ratio $e/m$ which is unrealistic, especially for classical electron models,
and requires the extreme RN solution as an exterior. Eq. (83) tells that
when the point $r=0$ is excluded, general CD solutions are possible. They
have the following characteristics 
\begin{equation}
z=\left( 1+\frac{r\nu ^{\prime }}2\right) ^{-2}\left( 1+C\frac{e^{-\nu }}{r^2%
}\right) ,  \label{htwentytwo}
\end{equation}
\begin{equation}
\kappa \rho =\left( 1+\frac{r\nu ^{\prime }}2\right) ^{-2}\left( 1+C\frac{%
e^{-\nu }}{r^2}\right) \frac{\nu ^{\prime }}r-\frac{z^{\prime }}r,
\label{htwentythree}
\end{equation}
\begin{equation}
\frac{q^2}{r^2}=1-\frac{1+r\nu ^{\prime }}{\left( 1+\frac{r\nu ^{\prime }}2%
\right) ^2}\left( 1+C\frac{e^{-\nu }}{r^2}\right) ,  \label{htwentyfour}
\end{equation}
\begin{equation}
\phi ^{\prime }=-e^{\nu /2}\frac{\left[ r^4\nu ^{\prime 2}-4C\left( 1+r\nu
^{\prime }\right) e^{-\nu }\right] ^{1/2}}{2r\left( r^2+Ce^{-\nu }\right)
^{1/2}}.  \label{htwentyfive}
\end{equation}
When $C=0$ they reduce to Eqs. (88)-(91). What is the physical significance
of such solutions? Let us consider a core of perfect fluid, occupying a ball
with radius $r_0$, with given $\nu $ and $p$. Then $z$ is determined from
Eq. (83) with $C=0$: 
\begin{equation}
z=\frac 1{\left( 1+\frac{r\nu ^{\prime }}2\right) ^2}\left[ 1+\frac{C\left(
r\right) }{r^2e^\nu }\right] ,  \label{htwentysix}
\end{equation}
\begin{equation}
C\left( r\right) =4\kappa \int_0^r\left( 1+\frac{r\nu ^{\prime }}2\right)
e^\nu r^3pdr.  \label{htwentyseven}
\end{equation}
At the junction $z$ is equivalent to Eq. (122) where $C=C_0\equiv C\left(
r_0\right) $. An observer cannot understand whether the interior solution
consists of perfect fluid or general CD since the only imprint left by the
pressure distribution inside is a constant. In principle, the RN metric
should be taken as an exterior, but there are no obstacles to take a general
CD metric and postpone the junction to RN till another point $r_1>r_0$. Thus
we obtain a triple-layered model with a perfect fluid core up to $r_0$ (zone
I), a halo of general CD from $r_0$ to $r_1$ (zone II) and the RN solution
for $r>r_1$ (zone III). The metric component $z$ is given correspondingly by
Eq. (126), Eq. (122) with $C_0$ and Eq. (44). For this purpose we choose a
continuous, together with its derivative, function $\nu $ in the region $%
0<r<r_1$. At the first junction $z$ is also continuous, while the pressure
drops to zero. The density is finite. In zone II the pressure remains zero,
while the density continues to decrease. Finally, at $r_1$ the density also
drops to zero (perhaps with a jump) and a RN solution follows till infinity,
making the composite solution asymptotically flat. As seen from Eqs.
(123),(124), the constant $C_0$ shifts $e/m$ from $1$, like a perfect fluid
solution, occupying zones I and II would do.

This picture will be backed by an explicit example with the ansatz (115).
Since the details should not depend on the form of $p\left( r\right) $ but
only on the value of $C$, we shall discuss first zone II and its junction
with zone III. In the interval $\left[ r_0,r_1\right] $ we have 
\begin{equation}
z=\frac{\left( a+br^2\right) \left[ C_0+r^2\left( a+br^2\right) \right] }{%
r^2\left( a+br^2\right) ^2},  \label{htwenteight}
\end{equation}
while in general 
\begin{equation}
\kappa \rho =\frac zr\left[ \frac 2r+\frac{8br}{a+2br^2}-\frac{C^{\prime
}\left( r\right) +2r\left( a+2br^2\right) }{C\left( r\right) +ar^2+br^4}%
\right] .  \label{htwentnine}
\end{equation}
For CD $C^{\prime }=0$ and one can easily show that the density is positive
for positive $C_0$. It decreases towards zero when $r\rightarrow \infty $.
The charge function is given by Eqs. (115), (124). Its square is positive
when 
\begin{equation}
C_0<\frac{b^2r_0^6}{a+3br_0^2}.  \label{hthirty}
\end{equation}
The total mass and charge are given by Eqs. (45),(46) 
\begin{equation}
m=r_1\left( 1-a-2br_1^2\right) ,  \label{hthirtyone}
\end{equation}
\begin{equation}
e^2=r_1^2\left( 1-a-3br_1^2\right) .  \label{hthirtytwo}
\end{equation}
The junction condition $-\lambda =\nu $ at $r_1$ gives 
\begin{equation}
C_0=m^2-e^2.  \label{hthirtythree}
\end{equation}
This relation clearly shows the effect of pressure in zone I on the
mass-charge ratio, not altered by the CD halo in zone II. The result in zone
I was found in Ref. \cite{six}.

The mass and squared charge are positive when $1-a>3br_1^2$. The constant $%
C_0$ is positive when 
\begin{equation}
4br_1^2\left( a+br_1^2\right) >a^2-a+br_1^2.  \label{hthirtyfour}
\end{equation}
A sufficient condition is 
\begin{equation}
\frac 14<br_1^2<\frac 13\left( 1-a\right) ,  \label{hthirtyfive}
\end{equation}
which requires also $a<0.24$. Let us take $a=0.01$ and $br_1^2=0.3$. Then
Eq. (130) is satisfied when $\left( \frac{r_0}{r_1}\right) ^2\geq 0.8$ and
utilizing the equality we get $r_1=4.013\sqrt{C_0}$, $r_0=3.588\sqrt{C_0}$, $%
b=0.075/\sqrt{C_0}$. This proves the existence of CD solutions with $\left|
e\right| /m\neq 1$. They sustain the value of $C$, obtained in zone I from a
perfect fluid solution with positive pressure.

It seems that there are no solutions of the type $\left( \nu ,p>0\right) $
in the literature. It must be stressed that this case is completely general,
unlike the cases of constant $T_0^0$ or with $\rho +p=0$, discussed in the
previous sections. Every perfect fluid solution may be reformulated as a $%
\left( \nu ,p\right) $ case. Let us proceed with the ansatz (115) which
transforms Eq. (127) into 
\begin{equation}
C\left( r\right) =4\kappa \int_0^r\left( a+2br^2\right) r^3pdr.
\label{hthirtysix}
\end{equation}
We base the discussion on $C\left( r\right) $ as a fundamental object,
hence, it is convenient to select a more involved $p$ leading to a simple $%
C\left( r\right) $. Let us take 
\begin{equation}
\kappa p=\frac{p_0-p_1r^2}{a+2br^2}  \label{hthirtyseven}
\end{equation}
and then 
\begin{equation}
C\left( r\right) =r^4\left( p_0-\frac 23p_1r^2\right) ,  \label{hthirtyeight}
\end{equation}
\begin{equation}
z=\frac{\left( a+br^2\right) ^2}{\left( a+2br^2\right) ^2}\left[ 1+\frac{%
r^2\left( p_0-\frac 23p_1r^2\right) }{a+br^2}\right] ,  \label{hthirtynine}
\end{equation}
\begin{equation}
\frac{q^2}{r^2}=\frac{r^4}{\left( a+2br^2\right) ^2}\left( b-bp_0-\frac 13%
ap_1\right) ,  \label{hforty}
\end{equation}
where $z$ follows from Eq. (126) while $q$ is calculated from the general
expression (84). The density is given by Eq. (129). Here $p_0$ and $p_1$ are
positive constants, connected by $p_0=p_1r_0^2$. The ansatz (115) yields $%
\nu _0=b/a$ and the condition $\rho \left( 0\right) >0$ becomes $%
8b/5a>\kappa p\left( 0\right) =p_0/a$. This is weaker than a necessary
condition for positive $q^2$, $b>p_0$. At the boundary the density is always
positive since $C^{\prime }\left( r_0\right) =0$ and Eq. (129) coincides
with the expression for general CD. We have also $p_0=3C_0/r_0^4$ and using
the numerical values of the CD example, one can express $p_0$ and $p_1$ in
terms of $C_0$, in addition to $b,r_0$ and $r_1$. It can be shown that for $%
C>0.06$, $\rho $ and $q^2$ are positive and the solution is physically
realistic. Thus we have constructed explicitly a triple solution of the type
PF-CD-RN. This ends the discussion of case $\left( \nu ,p\right) $.

The case $\left( \lambda ,p\right) $ relies on Eq. (26). Even the simplest
constant $z$ leads to special functions or non-integrable equations for $y$
when $p$ is chosen physically realistic.

\section{The case $\left( \nu ,q\right) $}

The cases $\left( \nu ,q\right) $ and $\left( \lambda ,q\right) $ comprise
the most direct generalizations of the numerous uncharged solutions. Setting 
$q=0$ one obtains one of them with the chosen ansatz for $\nu $ or $\lambda $%
. Like the previous section, the case $\left( \nu ,q\right) $ is much
simpler. The third main equation is Eq. (14). Now $f_1$ is not simply $%
-2g^{\prime }$ but it can be represented as a linear combination of $%
g^{\prime },g$ and $y$ which take care for $y^{\prime \prime },y^{\prime }$
and $y$ respectively. This is a general result which holds for any of Eqs.
(14)-(16), (19). In this particular case we have 
\begin{equation}
f_1=-2g^{\prime }+\frac{8g}r-4y,  \label{hfortyone}
\end{equation}
\begin{equation}
e^F=\frac{r^2e^{2h}}{e^\nu \left( 1+\frac{r\nu ^{\prime }}2\right) ^2},
\label{hfortytwo}
\end{equation}
\begin{equation}
H=2\int e^\nu \left( 1+\frac{r\nu ^{\prime }}2\right) e^{-2h}r^{-3}\left( 
\frac{2q^2}{r^2}-1\right) dr,  \label{hfortythree}
\end{equation}
\begin{equation}
h=\int \frac{\nu ^{\prime }dr}{1+\frac{r\nu ^{\prime }}2}  \label{hfortyfour}
\end{equation}
and $z$ is given by Eq. (22). We shall use an ansatz, slightly more general
than (115) 
\begin{equation}
e^\nu =\left( a+br^2\right) ^k,  \label{hfortyfive}
\end{equation}
where $k$ is an integer. In the uncharged case $k=1$ leads to Tolman's
solution IV \cite{fortytwo}. The case $k=2$ was discussed first by Wyman 
\cite{thirtyeight} as an extension of Tolman's solution VI and later was
studied in detail by Adler \cite{thirtynine}. Solution with $k=3$ was given
by Heintzmann \cite{fiftythree} (see also Ref. \cite{fiftytwo}). The general
class was studied by Korkina \cite{fiftyfour}, but her only explicit
solution was the one of Heintzmann. Later Durgapal \cite{fiftyfive} studied
in detail the cases $k=1-5$. All of them satisfy the physical criteria used
in this paper, but some have irregular behaviour of the speed of sound $%
dp/d\rho $ and $p/\rho $.

Going to the charged case, Eq. (145) and Eqs. (45), (46) give 
\begin{equation}
e^2=q^2\left( x_0\right) ,  \label{hfortysix}
\end{equation}
\begin{equation}
\frac m{r_0}=1+\frac{1+\left( k+1\right) \tau x_0}{1+\left( 2k+1\right) \tau
x_0}\left( \frac{e^2}{x_0}-1\right) ,  \label{hfortyseven}
\end{equation}
where $\tau =b/a$, $x=r^2$. Let us consider first the case $k=1$. Then 
\begin{equation}
z=\frac{1+\tau x}{1+2\tau x}\left( 1+Cx+2xQ\right) ,  \label{hfortyeight}
\end{equation}
\begin{equation}
Q=\int_0^x\frac{q^2}{x^3}dx.  \label{hfortynine}
\end{equation}
The pressure and the density are given by 
\begin{equation}
\kappa \rho =\frac{\tau -3C\left( 1+\tau x\right) }{1+2\tau x}+\frac{2\tau
\left( 1+Cx\right) }{\left( 1+2\tau x\right) ^2}-\frac{2\left( 1+\tau
x\right) }{1+2\tau x}\left( 3Q+\frac{2q^2}{x^2}\right) +\frac{4\tau xQ}{%
\left( 1+2\tau x\right) ^2},  \label{hfifty}
\end{equation}
\begin{equation}
\kappa p=\frac{\tau +C+3\tau Cx+2\left( 1+3\tau x\right) Q}{1+2\tau x}+\frac{%
q^2}{x^2}.  \label{hfiftyone}
\end{equation}
When $q=0$ these expressions coincide with the pressure and the density from
Refs. \cite{fortytwo,fiftyfive}. We have $\kappa \rho \left( 0\right)
=3\left( \tau -C\right) $. Let us choose $q^2=K^2x^3$. Then the condition $%
p\left( x_0\right) =0$ yields an expression for $C$%
\begin{equation}
C=-\frac{\tau +\left( 3+8\tau x_0\right) K^2x_0}{1+3\tau x_0},
\label{hfiftytwo}
\end{equation}
which means that $C$ is negative and consequently $\rho \left( 0\right) $ is
positive. The condition $z=e^\nu $ at $x_0$ defines $a$ as 
\begin{equation}
a=\frac{1+Cx_0+2K^2x_0^2}{1+2\tau x_0}.  \label{hfiftythree}
\end{equation}
Together with Eq. (146) which becomes $e^2=K^2x_0^3$ and Eq. (147) for $k=1$
we can express $c,a,e,m$ through $x_0,\tau ,K$.

Let us consider next the case $k=2$. Then Eqs. (142)-(145) give 
\begin{equation}
e^F=\frac x{\left( a+3bx\right) ^{2/3}},  \label{hfiftyfour}
\end{equation}
\begin{equation}
z=1+Ce^F+e^F\int \frac{2\left( a+bx\right) q^2dx}{\left( a+3bx\right)
^{1/3}x^3}.  \label{hfiftyfive}
\end{equation}
When $q=0$ this is the result of Adler who used the simplifying assumption $%
e^FH=1$ instead of the ansatz (145). When $q^2=K^2x^3$ we obtain 
\begin{equation}
z=1+\frac{Cx}{\left( a+3bx\right) ^{2/3}}+\frac{4a}{5b}K^2x+\frac 25K^2x^2.
\label{hfiftysix}
\end{equation}
This is the corrected result of Nduka \cite{thirtysix} who wrote the field
equations with an error in the sign of $q^2$. In all his results $K^2$
should be replaced by $-K^2$ and the conclusions correspondingly altered.

In another paper Singh and Yadav \cite{thirtyseven} simplified Eqs. (14),
(21) by demanding $f_1=-f_0$. This leads to the following equation for $y$%
\begin{equation}
r^2y^{\prime \prime }-ry^{\prime }-\frac{2q^2}{r^2}y=0.  \label{hfiftyseven}
\end{equation}
Letting $q=Kr$ they obtain the Euler equation. It has three types of
solutions, all of which are singular at $r=0$ since $y\sim r^{a_1}$. If $K=0$
the non-singular solution of Adler is again recovered.

\section{The case $\left( \nu ,n\right) $}

This is the case with given $\nu $ and the perfect fluid satisfying the
linear equation of state Eq. (17). Realistic solutions have $0<n\leq 1$
which means constant speed of sound and causal behavior. Surprisingly, the
charged case is much easier than the uncharged one. The third main equation
is Eq. (19) with $n\neq -1$. An expression for $f_1$, analogous to Eq.
(141), may be written 
\begin{equation}
f_1=-2g^{\prime }+\frac{16ng}{\left( n+1\right) r}-\frac{8n\left(
9n+1\right) }{\left( n+1\right) ^2}y.  \label{hfiftyeight}
\end{equation}
Then $F$ and $H$ read 
\begin{equation}
e^F=\frac{r^{-\left( \alpha +1\right) }e^{\beta h_1}}{e^\nu \left( A+\frac{%
r\nu ^{\prime }}2\right) ^2},  \label{hfiftynine}
\end{equation}
\begin{equation}
H=2\int e^\nu \left( A+\frac{r\nu ^{\prime }}2\right) r^\alpha e^{-\beta
h_1}\left( 1-\frac{2\kappa p_0}{n+1}r^2\right) dr,  \label{hsixty}
\end{equation}
where 
\begin{equation}
h_1=\int \frac{\nu ^{\prime }dr}{A+\frac{r\nu ^{\prime }}2},
\label{hsixtyone}
\end{equation}
\begin{equation}
\alpha =\frac{1-3n}{5n+1},  \label{hsixtytwo}
\end{equation}
\begin{equation}
A=\frac{5n+1}{n+1},  \label{hsixtythree}
\end{equation}

\begin{equation}
\beta =\frac{4n\left( 9n+1\right) }{\left( n+1\right) \left( 5n+1\right) }.
\label{hsixtyfour}
\end{equation}
We shall study mainly the range $0\leq n\leq \infty $ and occasionally some
negative $n$. The case $n=-1/5$ demands special treatment to be done at the
end of the section. The case $n=-1$ was discussed in Sec. IV. Each of the
coefficients $\alpha ,\beta $ and $A$ has the same limits when $n\rightarrow
\pm \infty $. When $0\leq n\leq \infty $ their ranges are $-3/5\leq \alpha
\leq 1$, $1\leq A\leq 5$, $0\leq \beta \leq 36/5$. Expressions (159), (160)
are generic for the cases $\left( \nu ,q\right) ,\left( \nu ,n\right)
,\left( \nu ,\rho \right) $. The case $\left( \nu ,q\right) $, discussed in
the previous section, is obtained by putting $\alpha =-3$, $\beta =2$, $A=1$%
, changing the sign of $H$ and an obvious replacement for $q$ instead of $%
p_0 $. The case $\left( \nu ,\rho \right) $ will be studied in the next
section.

In the present case Eq. (162) shows that $\alpha +1$ is always positive for $%
n\in \left[ 0,\infty \right] $. Therefore, $z$ will have a pole unless $C=0$%
. Hence 
\begin{equation}
z=e^FH.  \label{hsixtyfive}
\end{equation}
After $z$ is found, Eqs. (9),(10),(17) allow to extract the fluid components
from the metric 
\begin{equation}
\kappa \rho =\frac 1{n+1}\left[ \kappa p_0+\frac zr\left( \nu ^{\prime }-%
\frac{z^{\prime }}z\right) \right] ,  \label{hsixtysix}
\end{equation}
\begin{equation}
\kappa p=\frac 1{n+1}\left[ \frac{nz}r\left( \nu ^{\prime }-\frac{z^{\prime }%
}z\right) -\kappa p_0\right] ,  \label{hsixtyseven}
\end{equation}
\begin{equation}
\frac{q^2}{r^2}=1-z-\frac{r^2}{n+1}\left[ \kappa p_0+\frac zr\left( \nu
^{\prime }+n\frac{z^{\prime }}z\right) \right] .  \label{hsixtyeight}
\end{equation}
Since $p_0>0$, a necessary condition for positive pressure is $\nu ^{\prime
}>z^{\prime }/z$ and this leads to $\rho >\frac{p_0}{n+1}$. It is more
convenient to express $p$ and $q$ from Eqs. (17), (84) respectively and lay
all the difficulty in the calculations on the density. Plugging Eqs. (159),
(160), (165) into Eq. (166) we obtain 
\begin{eqnarray}
\left( n+1\right) \kappa \rho &=&\kappa p_0+\frac{4\kappa p_0}{\left(
n+1\right) \left( A+\frac{r\nu ^{\prime }}2\right) }+\frac 2{r^2}\left[ 
\frac zA-\frac 1{A+\frac{r\nu ^{\prime }}2}\right] +\left( 2-\frac \beta {A+%
\frac{r\nu ^{\prime }}2}\right) \frac{z\nu ^{\prime }}r+  \nonumber \\
&&\frac{z\left( r\nu ^{\prime }\right) ^{\prime }}{r\left( A+\frac{r\nu
^{\prime }}2\right) },  \label{hsixtynine}
\end{eqnarray}
which is an analog of Eq. (86) and leads to similar conclusions for $\nu $.
The poles in the two terms in the square brackets cancel when $r\rightarrow
0 $ because $z\rightarrow 1$ and $\nu ^{\prime }\rightarrow 2\nu _0r$. This
condition makes $\rho $ a well-defined function. We shall use for $\nu $ an
ansatz which generalizes those in Eqs. (115), (145) 
\begin{equation}
e^\nu =\left( a+br^s\right) ^k,  \label{hseventy}
\end{equation}
where $s\geq 2$ is not necessarily an integer. We obtain for $z$%
\begin{equation}
z=\frac{4\int \left( a+bx\right) ^{k-1}\left[ 2aA+\left( 2A+ks\right)
bx\right] ^\gamma \left( 1-\frac{2\kappa p_0}{n+1}x^{2/s}\right) x^{\frac{%
\alpha +1}s-1}dx}{sx^{\frac{\alpha +1}s}\left( a+bx\right) ^{k-2}\left[
2aA+\left( 2A+ks\right) bx\right] ^{1+\gamma }},  \label{hseventyone}
\end{equation}
\begin{equation}
\gamma =1-\frac{2k\beta }{2A+ks}.  \label{hseventytwo}
\end{equation}
There are several ways to simplify the expression under the integral. One of
them is to set $\gamma $ (which is not an integer, in general) equal to zero
for any $n$ choosing $k$ and $s$ appropriately. From Eq. (172) the following
relation arises 
\begin{equation}
\frac s2=\frac{\left( 36k-25\right) n^2+\left( 4k-10\right) n-1}{k\left(
n+1\right) \left( 5n+1\right) }.  \label{hseventythree}
\end{equation}
The R.H.S. should be $\geq 1$ which yields the condition $n\geq n_k$%
\begin{equation}
n_k=\frac{k+5+4\sqrt{k\left( 2k+1\right) }}{31k-25}.  \label{hseventyfour}
\end{equation}
The first few values are $n_1\approx 2.15,$ $n_2\approx 0.53,$ $n_3\approx
0.39,$ $n_4=1/3$. The first is beyond the physical range, the fourth is an
exact number. One cannot probe the region $n<n_\infty \approx 0.21$ with
this method.

Let us replace now $z$ from Eq. (171) (with $\gamma =0$) into Eq. (169) and
calculate $\rho \left( 0\right) $. When $s>2$ the last two terms do not
contribute, while from $z$ only the term proportional to $a^{k-1}p_0$
counts. The result is very simple 
\begin{equation}
\rho \left( 0\right) =\frac{3p_0}{3n+1}.  \label{hseventyfive}
\end{equation}
The equation of state at the centre reads $\rho \left( 0\right) +3p\left(
0\right) =0$ which is exactly the equation of state for the ESU \cite
{fifty,fiftytwo}. The pressure is negative at the centre, which is
unrealistic.

In this connection it is worth to point out another relation to ESU even
when $\gamma $ does not vanish. Let us take in $\nu $ and $z$ the limit $b=0$%
. Then 
\begin{equation}
e^\nu =a^k,  \label{hseventysix}
\end{equation}
\begin{equation}
z=1-\frac{\kappa p_0}{3n+1}r^2,  \label{hseventyseven}
\end{equation}
$\rho $ is given by Eq. (175) and $q=0$. Now we obtain ESU for any $r$
independently of $s,k,n$ and $p_0$. Hence, the results in this section may
be considered as a specific 'electrification' of ESU like the generalization
of de Sitter solution in Sec.IV and of flat spacetime in Sec.V.

Going back to the case $\gamma =0$, we reach the conclusion that $s=2$ is a
necessary condition for realistic solutions. The overlapping bands of
solutions with $n\geq n_k$ collapse to an infinite set of discrete values $%
n_k$, $k\geq 2$. Then the ansatz (170) coincides with (145) and we shall use
the corresponding notation. A series of models is obtained, parameterized by 
$k$ and $n$ is given by Eq.(174). They have 
\begin{equation}
e^\nu =a^k\left( 1+t\right) ^k,  \label{hseventyeight}
\end{equation}
\begin{equation}
z=\frac{\int \left( 1+t\right) ^{k-1}t^{\frac{\alpha +1}2-1}\left( 1-\frac 2{%
n+1}\mu t\right) dt}{t^{\frac{\alpha +1}2}\left( 1+t\right) ^{k-2}\left[
A+\left( A+k\right) t\right] },  \label{hseventynine}
\end{equation}
where $\mu =\kappa p_0/\tau $. Eq. (169) becomes 
\begin{eqnarray}
\left( n+1\right) \frac{\kappa \rho }\tau &=&\mu +\frac{4\mu \left(
1+t\right) }{\left( n+1\right) \left[ A+\left( A+k\right) t\right] }+\frac 2t%
\left[ \frac zA-\frac{1+t}{A+\left( A+k\right) t}\right] +  \label{heighty}
\\
&&\frac{2kz}{1+t}\left[ 2-\beta \frac{1+t}{A+\left( A+k\right) t}\right] +%
\frac{4kz}{\left( 1+t\right) \left[ A+\left( A+k\right) t\right] }. 
\nonumber
\end{eqnarray}
Now additional terms contribute to $\rho \left( 0\right) $ and $p\left(
0\right) $ and they can be made positive by a proper choice of $\tau $ and $%
p_0$. The pressure is given by Eq. (17) and the charge function by Eq. (84)
which becomes 
\begin{equation}
\tau q^2=t\left[ 1-\frac{1+\left( 2k+1\right) t}{1+t}z+\frac{\kappa p}\tau
t\right] .  \label{heightyone}
\end{equation}

Let us discuss the particular case $k=2$, $n_2=0.53$. Then 
\begin{equation}
z=\frac{1.19+0.35t-0.46\mu t-0.27\mu t^2}{1.19+2.19t}  \label{heightytwo}
\end{equation}
and it can be shown that $p\left( 0\right) >0$ when $\mu <6.34$. Let us take 
$\mu =1$. Then 
\begin{equation}
\frac{\kappa p}\tau =\frac{-4.05-6.1t+\frac{4z}{1+t}\left( 4.57+6.57t\right) 
}{6.87+12.64t},  \label{heightythree}
\end{equation}
\begin{equation}
\tau q^2=t\left( 1-\frac{1+5t}{1+t}z+\frac{\kappa p}\tau t\right) .
\label{heightyfour}
\end{equation}
The pressure is monotonously decreasing and vanishes at $t_0=0.66$. The
charge function reaches a maximum before $t_0$ and is still positive at $t_0$%
. This indicates that $\sigma $ from positive turns to negative. As in Eq.
(153) the condition $z\left( r_0\right) =e^\nu \left( r_0\right) $
determines $a$. The charge and the mass are determined by Eqs. (45),(46).
All constants may be written as functions of $r_0$: $a=0.37$, $b=0.24/r_0^2$%
, $m=0.32r_0$, $e=0.49r_0$. The ratio $e/m>1$.

Another interesting case is $k=4$ with $n_4=1/3$ exactly. When $p_0=0$ and $%
q=0$ this is the equation of state of pure incoherent radiation and the case
is not integrable \cite{fortysix}. This specific charged version, which is
not the only possible one, however, is integrable and assuming $\mu =2/3$ we
have 
\begin{equation}
z=\frac{1+\frac 23t-\frac 27t^3-\frac 19t^4}{\left( 1+t\right) ^2\left(
1+3t\right) },  \label{heightyfive}
\end{equation}
\begin{equation}
\frac{\kappa p}\tau =\frac{5+9t}{12\left( 1+3t\right) }+\frac{3\left(
3+7t\right) z}{2\left( 1+t\right) \left( 1+3t\right) }-\frac{\frac 73+3t+%
\frac 97t^2+\frac 19t^3}{4\left( 1+t\right) ^2\left( 1+3t\right) }-\frac 23,
\label{heightysix}
\end{equation}
\begin{equation}
\tau q^2=t\left( 1-\frac{1+9t}{1+t}z+\frac{\kappa p}\tau t\right) .
\label{heightyseven}
\end{equation}
The pressure is positive, monotonously decreasing and vanishes at $t_0=0.54$%
. The charge density has a negative part, like the previous case. The
junction conditions yield $a=0.3$, $b=0.16/r_0^2,$ $m=0.5r_0$, $e=0.45r_0$.
This time $e/m<1$. Probably, all models in the series have a range of $\mu $
where they are physically realistic.

The method used to simplify Eq. (160), after all, resembles the
Korkina-Durgapal method, which was extended to the charged case $\left( \nu
,q\right) $ in the previous section. With $s=2$ the ansatz (170) is
equivalent to (145) and, effectively, the non-integer $\gamma $ was
exchanged for an integer $k$ in Eq. (171), which allows integration in
rational functions. There are other ways to simplify Eq. (171), even when $%
k=1$. One of them is to take again $n=1/3$. Then $\alpha =0,$ $A=2$, $\beta
=3/2,$ $\gamma =1/2$. We have 
\begin{equation}
e^\nu =a\left( 1+t\right) ,  \label{heightyeight}
\end{equation}
\begin{equation}
z=\frac{\left( 1+t\right) \int \left( 2+3t\right) ^{1/2}\left( 1-\frac 32\mu
t\right) t^{-1/2}dt}{t^{1/2}\left( 2+3t\right) ^{3/2}}.  \label{heightynine}
\end{equation}
The integral is expressed in elementary functions 
\begin{equation}
z=\frac{1+t}{t^{1/2}\left( 2+3t\right) ^{3/2}}\left[ -\frac 14\left( \mu
-4+3\mu t\right) \sqrt{\left( 2+3t\right) t}+\frac{4+\mu }{12}\sqrt{3}\ln
\left( 3t+1+\sqrt{3t\left( 2+3t\right) }\right) \right] ,  \label{hninety}
\end{equation}
where the integration constant was chosen so that $z\left( 0\right) =1$. The
pressure and the charge are given by Eq. (17), (169) and (181) 
\begin{equation}
\frac{\kappa p}\tau =-\frac{3\left( 1+2t\right) }{4\left( 2+3t\right) }\mu +%
\frac{9z}{4\left( 2+3t\right) }+\frac 1{2t}\left( \frac z2-\frac{1+t}{2+3t}%
\right) ,  \label{hninetyone}
\end{equation}
\begin{equation}
\tau q^2=t\left( 1-\frac{1+3t}{1+t}z\right) +\frac{\kappa p}\tau t^2.
\label{hninetytwo}
\end{equation}
The pressure is well-behaved for any $\mu $ and vanishes at $t_0$. However, $%
q^2$ also has a zero, but at $t_1$, and becomes negative in a region where $%
t>t_1$. Unfortunately, for most values of $\mu $ the inequality $t_1<t_0$
holds, meaning that $q^2<0$ in some part of the interior. This is grossly
unrealistic. We have been unable to find by computer simulations any
realistic solution.

Another case with $k=1$ which leads to simple functions can be guessed from
Eq. (164). When $n=-1/9$, $\beta \equiv 0$. We have also $\alpha =3$, $A=1/2$%
, $\gamma =1$. Although the pressure is negative, it is worth being
discussed. Eqs. (188), (192) still hold. The integral in Eq. (171) is
elementary and gives 
\begin{equation}
z=\frac{1+t}{\left( 1+3t\right) ^2}\left[ 1+\left( 2-\frac 32\mu \right) t-%
\frac{27}8\mu t^2\right] .  \label{hninetythree}
\end{equation}
Eq. (169) leads to $\rho $ and $p$ similar in form, since Eq. (17) is
satisfied. Like usual, we give only the formula for the pressure 
\begin{equation}
\frac{\kappa p}\tau =\frac{1+t}{2\left( 1+3t\right) ^2}\left[ 2+\frac 32\mu
+\left( 3+\frac{27}8\mu \right) t\right] -\frac{9\left( 1+2t\right) }{%
4\left( 1+3t\right) }\mu -\frac{3z}{2\left( 1+3t\right) }.
\label{hninetyfour}
\end{equation}
The values at the centre are 
\begin{equation}
\frac{\kappa p}\tau \left( 0\right) =-\frac 12\left( 1+3\mu \right) M\text{, 
}\frac{\kappa \rho }\tau \left( 0\right) =\frac 92\left( 1+\mu \right) .
\label{hninetyfive}
\end{equation}
On the other hand 
\begin{equation}
\frac{\kappa p}\tau \left( \infty \right) =\frac 16-\frac 98\mu .
\label{hninetysix}
\end{equation}
Thus, when $0<\mu <4/27$ the pressure is monotonously increasing, crosses
the $t$ axis and forms a boundary of fluid with tension inside. Let us take
the $\gamma $-law with $\mu =0$. The charge function $q^2$ is positive and
increasing. The junction conditions give $a=0.55,$ $b=0.11/r_0^2,$ $%
m=0.23r_0 $, $e=0.35r_0$ where $t_0=0.2$. This solution resembles the
classical models of the electron in Sec.IV.

Finally, some words should be said about the case $n=-1/5$, which is really
very special. Eqs. (159), (160) are replaced by 
\begin{equation}
e^F=\frac{4e^{-2h_2}}{r^6e^\nu \left( r\nu ^{\prime }\right) ^2},
\label{hninetyseven}
\end{equation}
\begin{equation}
H=\frac 12\int e^\nu \nu ^{\prime }r^6\left( 2-5\kappa p_0r^2\right)
e^{2h_2}dr,  \label{hninetyeight}
\end{equation}
\begin{equation}
h_2=2\int \frac{dr}{r^2\nu ^{\prime }}.  \label{hninetynine}
\end{equation}
Let us choose the simple ansatz (115) or (188) for $\nu $, written as 
\begin{equation}
e^\nu =a+bx=a\left( 1+\tau x\right) .  \label{th}
\end{equation}
Then Eqs. (197), (198) become 
\begin{equation}
e^F=\frac{a+bx}{b^2x^6}e^{1/\tau x},  \label{thone}
\end{equation}
\begin{equation}
H=-\frac b2\int \left( \frac 2{X^6}-\frac{5\kappa p_0}{X^7}\right)
e^{-X/\tau }dX,  \label{thtwo}
\end{equation}
where $X=1/x$. The integrals can be expressed through the $%
\mathop{\rm Ei}
\left( X\right) $ function. For example 
\begin{equation}
\int X^{-6}e^{-X/\tau }dX=e^{-1/\tau x}\sum_{k=1}^5\frac{\left( -\tau
\right) ^{-k+1}x^{6-k}}{\left( 6-1\right) ...\left( 6-k\right) }+\frac{%
\left( -\tau \right) ^{-5}}{5!}%
\mathop{\rm Ei}
\left( -\frac 1{\tau x}\right) .  \label{ththree}
\end{equation}
Then $z$, which is given again by Eq. (165), will have poles for $x=0$, not
compensated by $%
\mathop{\rm Ei}
$. Thus, already $z$ is ill-defined, although the case is formally soluble.
In addition $n<0$. When $p_0=0$ and $q=0$, this case is soluble \cite
{fortysix}. It is connected by a Buchdahl transformation to the de Sitter
solution with $n=-1$ \cite{fiftysix,fiftyseven} and is best expressed in
isotropic coordinates. This transformation, however, does not work for
electrostatic fields and probably is not applicable to the present case of
perfect fluid plus an electrostatic field. Nevertheless, some memories from
the uncharged case seem to have been preserved in a rather twisted form in
the presence of charge.

The only paper which comes near to the results of the present section seems
to be Ref. \cite{thirtysix} which belongs to the $\left( \nu ,q\right) $
case and is mentioned in Sec.VI (see Eq. (156)). When $m/r_0$ is small, this
solution approximately satisfies Eq. (17) with $n=3/10$, value rather close
to $1/3$. The connection to our solutions is not clear.

A final remark is in order. It may be tempting to take the limit $%
e\rightarrow 0$ in one of the solutions obtained here and try to find
analytic expressions for the non-integrable at present cases of $\gamma $%
-law uncharged solutions \cite{fortysix}. However, for the ansatz (145) used
in this section, this leads to $r_0\rightarrow 0$ and to flat spacetime.
These solutions do not allow point-like models and this seems to be property
only of the CD solutions, discussed in the introduction. In the case of
general $\nu $, it seems that the above limit will produce charged solutions
with zero total charge. This is suggested by the behavior of $\sigma $ in
some of the models, namely, its change of sign in the interior. The proper
non-charged limit seems to be $q\left( r\right) \equiv 0$. Then, however,
Eq. (9) relates $\rho $ to $\lambda $ and we obtain the field equations of
the uncharged fluid. Hence, nothing is gained in comparison with the three
approaches described in Ref.\cite{fortysix}. They lead to Abel differential
equations of the second kind with few integrable cases.

\section{The case $\left( \nu ,\rho \right) $}

This case arises as a special limit of $\left( \nu ,n\right) $ when $%
n\rightarrow \infty ,$ $p_0\rightarrow \infty $. One sees from Eq. (17) that
in this limit $\rho \rightarrow \rho _0=p_0/n$ which can be made a finite
constant. More generally, the third main equation (15) is obtained from Eq.
(19) plus the change $\frac{p_0}{n+1}\rightarrow \rho \left( r\right) $.
Eqs. (162)-(164) give $\alpha =-3/5,$ $A=5$, $\beta =36/5$. The constant $C$
in Eq. (22) is zero again to prevent a pole at the centre. Eqs. (158)-(161),
(165) yield 
\begin{equation}
f_1=-2g^{\prime }+\frac{16g}r-72y,  \label{thfour}
\end{equation}
\begin{equation}
z=\frac{2e^{\frac{36}5h_3}}{r^{2/5}e^\nu \left( 5+\frac{r\nu ^{\prime }}2%
\right) ^2}\int e^\nu \left( 5+\frac{r\nu ^{\prime }}2\right) r^{-3/5}e^{-%
\frac{36}5h_3}\left( 1-2\kappa \rho r^2\right) dr,  \label{thfive}
\end{equation}
\begin{equation}
h_3=\int \frac{\nu ^{\prime }dr}{5+\frac{r\nu ^{\prime }}2}.  \label{thsix}
\end{equation}
The parameters $n$ and $p_0$ have been exchanged effectively for the density
function and in this respect Eq. (205) resembles Eqs. (142),(143) from the $%
\left( \nu ,q\right) $ case. When $\rho $ is constant we obtain a
generalization of the Schwarzschild interior solution \cite{fiftyeight}. It
is much more complicated than the other generalization with constant $T_0^0$%
, which imposes just the simple ansatz (50) on the metric and was discussed
in Sec.III. As seen from Eq. (205), the case $\rho =0$ is non-trivial and
leads to purely electromagnetic mass models, because the mass function and $%
T_0^0$ receive contribution only from the charge. Formally, it is an analog
of the $\gamma $-law equation of state ($p_0=0$) from the previous section.

Let us use again the ansatz (170) for $\nu $. Now Eq. (172) reads 
\begin{equation}
\gamma =1-\frac{72k}{50+5ks},  \label{thseven}
\end{equation}
which is nullified when 
\begin{equation}
s=\frac{72k-50}{5k}.  \label{theight}
\end{equation}
This expression always gives $s\geq 2$ for $k\geq 1$. The analog of Eq.
(171) with $\gamma =0$ is 
\begin{equation}
z=\frac{4\int \left( a+br^s\right) ^{k-1}\left( 1-2\kappa \rho r^2\right)
r^{-3/5}dr}{r^{2/5}\left( a+br^s\right) ^{k-2}\left[ 10a+\left( 10+ks\right)
br^s\right] }.  \label{thnine}
\end{equation}
Now $\rho $ is given and $p$ is found from Eq. (10) 
\begin{equation}
\kappa p=\frac zr\left( \nu ^{\prime }-\frac{z^{\prime }}z\right) -\kappa
\rho .  \label{thten}
\end{equation}
Let us calculate $p\left( 0\right) $. When $s>2$, just two terms contribute
and the result is simple 
\begin{equation}
p\left( 0\right) =-\frac 15\rho \left( 0\right) .  \label{theleven}
\end{equation}
This is similar to the relation following from Eq. (175). The pressure and
the density cannot be both positive at the centre, hence, we must take $s=2$%
. Then Eq. (208) shows that $k$ is not an integer and we gain nothing,
trying to simplify $z$. The method which lead to an infinite series of
realistic solutions in the $\left( \nu ,n\right) $ case seems rather useless
here. Therefore, let us put $k=1$ in Eq. (171). Then $\gamma =-1/5$ and it
becomes 
\begin{equation}
z=\frac{1+t}{t^{1/5}\left( 5+6t\right) ^{4/5}}\int \frac{\left( 1-\frac{%
2\kappa \rho }\tau t\right) dt}{t^{4/5}\left( 5+6t\right) ^{1/5}}.
\label{thtwelve}
\end{equation}
When $\rho =\rho _0$ the integrals bring forth the hypergeometric function 
\begin{equation}
z=\frac{1+t}{\left( 1+\frac 65t\right) ^{4/5}}\left[ F\left( \frac 15,\frac 1%
5,\frac 65;-\frac{6t}5\right) -\frac{\kappa \rho _0}{3\tau }tF\left( \frac 15%
,\frac 65,\frac{11}5;-\frac{6t}5\right) \right] .  \label{ththirteen}
\end{equation}
The charge is given by Eq. (192), while the pressure follows from Eq. (210)
written as 
\begin{equation}
\frac{\kappa p}\tau =\frac{2z}{1+t}-2z_t-\frac{\kappa \rho }\tau .
\label{thfourteen}
\end{equation}

Searching for an elementary solution, one may consider some sophisticated
density profile, leading to a simple integral in Eq. (212). For example 
\begin{equation}
\frac{\kappa \rho }\tau =\frac 1{2t}\left[ 1-\left( 1+\frac 65t\right)
^{1/5}\left( 1-b_1t+b_2t^2\right) \right]  \label{thfifteen}
\end{equation}
leads to 
\begin{equation}
z=\frac{1+t}{\left( 1+\frac 65t\right) ^{4/5}}\left( 1-\frac{b_1}6t+\frac{b_2%
}{11}t^2\right) ,  \label{thsixteen}
\end{equation}
where $b_1,b_2$ are positive constants. We have $\rho \left( 0\right) >0$
when $b_1>6/25$ and $\rho $ is decreasing when $b_2>\frac{6b_1}{25}+\frac{72%
}{625}$. Taking $b_1=5$, $b_2=2$ a solution with realistic $z,\rho $ and $p$
is obtained, having a boundary at $t_0=0.52$. Unfortunately, $q^2<0$ in the
interior, although for larger $t$ it becomes positive. Perhaps, further fine
tuning of the density may lead to physically acceptable solutions.

\section{The cases $\left( \rho ,q\right) $, $\left( \lambda ,\rho \right) $
and $\left( \lambda ,q\right) $}

Having in mind the results in the previous section, it is no wonder that the
case $\left( \rho ,q\right) $, where there is control over the evasive
charge function, is much more popular in the literature. Eq. (9) easily
relates this case to $\left( \lambda ,\rho \right) $ and $\left( \lambda
,q\right) $, and Eq. (24) or (25) may be used as a third main equation. Eqs.
(7),(8) define $z$ in terms of $\rho $ and $q$%
\begin{equation}
z=1-\frac \kappa r\int_0^r\rho r^2dr-\frac 1r\int_0^r\frac{q^2}{r^2}dr.
\label{thseventeen}
\end{equation}
Eq. (9) shows that $\rho $ is regular when $z-1\sim r^2$ for small $r$. We
accept the ansatz 
\begin{equation}
z=1-ar^2-br^4.  \label{theighteen}
\end{equation}
It follows from Eq. (217) when $q=Kr^l$ with $l=2,3$ and 
\begin{equation}
\rho =\rho _0-\rho _1r^2,  \label{thnineteen}
\end{equation}
where $\rho _0,\rho _1$ are non-negative constants.. When $l=2$ or $l=3$ we
have respectively 
\begin{equation}
z=1-\frac 13\left( \kappa \rho _0+K^2\right) x+\frac{\kappa \rho _1}5x^2,
\label{thtwenty}
\end{equation}
\begin{equation}
z=1-\frac{\kappa \rho _0}3x+\frac 15\left( \kappa \rho _1-K^2\right) x^2.
\label{thtwentyone}
\end{equation}
The total charge and mass are given by 
\begin{equation}
e=Kx_0^{3/2},  \label{thtwentytwo}
\end{equation}
\begin{equation}
m=\frac 12\left[ a+\left( b+K^2\right) x_0\right] x_0^{3/2}.
\label{thtwentythree}
\end{equation}
When $\rho _1\neq 0$ this is a model of a gaseous sphere, both density and
pressure vanish at the boundary \cite{twentyfive,twentysix}. We have $\rho
_0=\rho _1x_0$. When $\rho _1=0$, $\rho _0\neq 0$ this is a generalization
of the incompressible fluid sphere for $l=3$ \cite{twentytwo} or $l=2$ \cite
{twentythree}. Finally, when $\rho _1=\rho _0=0$ we come to a model with
electromagnetic mass \cite{six}. The case $b\neq 0$ appears to be simpler,
due to a peculiarity in Eq. (24) and leads to elementary functions. Eq. (50)
demonstrates that the other case, $b=0$, belongs to the family of solutions
with constant $T_0^0$. Eq. (24) becomes in terms of $x$%
\begin{equation}
\left( 1-ax-bx^2\right) y_{xx}-\left( bx+\frac a2\right) y_x-\left( \frac b4+%
\frac{q^2}{2x^3}\right) y=0.  \label{thtwentyfour}
\end{equation}
The coefficient before $y$ is constant when $l=3$ (or $K=0$). Let us study
this case first. Eq. (221) gives 
\begin{equation}
d\equiv -\left( \frac b4+\frac{q^2}{2x^3}\right) =\frac 1{20}\left( \kappa
\rho _1-11K^2\right) .  \label{thtwentyfive}
\end{equation}
A change of variables brings Eq. (224) to (see Eq.2.1.164 from Ref.\cite
{fortynine}) 
\begin{equation}
y_{\xi \xi }+dy=0,  \label{thtwentysix}
\end{equation}
\begin{equation}
\xi =\int_0^x\frac{dx}{\sqrt{1-ax-bx^2}}.  \label{thtwentyseven}
\end{equation}
This integral has several different expressions according to the signs of $b$
and $a^2+4b$, but many properties of the solutions depend only on the facts
that $\xi \left( 0\right) =0$, $\xi $ and $\xi _x$ are positive. Eq. (226)
is easily solved and the last function to be determined, the pressure, is
given by Eq. (10), suitably rewritten as 
\begin{equation}
\kappa p=4\sqrt{z}\frac{y_\xi }y-2z_x-\kappa \rho .  \label{thtwentyeight}
\end{equation}
At the junction 
\begin{equation}
y\left( x_0\right) =\sqrt{z\left( x_0\right) }.  \label{thtwentynine}
\end{equation}
Then from Eq. (228) follows that the pressure vanishes when 
\begin{equation}
2y_\xi \left( x_0\right) =z_x\left( x_0\right)  \label{ththirty}
\end{equation}
is satisfied. The charge is given by Eq. (222) while the mass in Eq. (223)
becomes 
\begin{equation}
m=\frac 12\left[ \frac{\kappa \rho _0}3+\left( 6K^2-\kappa \rho _1\right) 
\frac{x_0}5\right] x_0^{3/2}.  \label{ththirtyone}
\end{equation}
When $\rho _1=0$ the mass is positive. When $\rho _1\neq 0$ this assertion
still holds because then $\rho _1x_0=\rho _0$. The positivity of the mass
does not impose any conditions on the constants $\rho _0$, $x_0$ and $K$
which determine the solution.

Inserting $z$ and $\rho $ in Eq. (228) yields 
\begin{equation}
\kappa p=4\sqrt{z}\frac{y_\xi }y-\frac 13\kappa \rho _0+\frac 15\left(
3\kappa \rho _1+2K^2\right) x.  \label{ththirtytwo}
\end{equation}
Then 
\begin{equation}
\kappa p\left( 0\right) =4\frac{y_\xi }y\left( 0\right) -\frac 13\kappa \rho
_0  \label{ththirtythree}
\end{equation}
should be positive. To be more concrete, let us introduce the three classes
of solutions of Eq. (226) for $d=0$, $d<0$ and $d>0$ respectively: 
\begin{equation}
y=C_1+C_2\xi ,  \label{ththirtyfour}
\end{equation}
\begin{equation}
y=C_1e^{-\sqrt{-d}\xi }+C_2e^{\sqrt{-d}\xi },  \label{ththirtyfive}
\end{equation}
\begin{equation}
y=C_1\sin \sqrt{d}\xi +C_2\cos \sqrt{d}\xi .  \label{ththirtysix}
\end{equation}
The condition $y>0$ and Eq. (233) lead to $C_1>0$, $C_2>0$ when $d\geq 0$
and to $C_2>0$, $C_2>C_1$ when $d<0$. Therefore, $e^\nu $ increases with $r$
until it meets $z$ which decreases. We have $y\left( 0\right) <1$.

1) Case $d=0$. Eq. (225) means $\kappa \rho _1=11K^2$, while Eq. (221) gives 
$b=-2K^2$. We also have 
\begin{equation}
x_0=\frac{\kappa \rho _0}{11K^2}  \label{ththirtyseven}
\end{equation}
and the solution is determined by two constants, $\rho _0$ and $K$. Eqs.
(229),(230) result in 
\begin{equation}
C_2=\frac 1{66}\kappa \rho _0,  \label{ththirtyeight}
\end{equation}
\begin{equation}
C_1=-\frac 1{66}\kappa \rho _0\xi _0+\sqrt{z_0},  \label{ththirtynine}
\end{equation}
\begin{equation}
z_0=1-\frac 53v^2\text{, }v\equiv \frac{\kappa \rho _0}{11\left| K\right| }
\label{thforty}
\end{equation}
and $y$ is given by Eq. (234). $C_1$ is real when $v\leq \sqrt{3/5}$. The
charge is given by Eq. (222) and the mass is 
\begin{equation}
m=\frac 4{33}\kappa \rho _0x_0^{3/2},  \label{thfortyone}
\end{equation}
leading to the ratio $\left| e\right| /m=3/4v\geq \sqrt{15}/4.$ Due to the
negative $b$ Eq.(227) reads 
\begin{equation}
\xi =\frac 1{\sqrt{-b}}\ln \left| -\frac{a+2bx}{2\sqrt{-b}}+\sqrt{z}\right| -%
\frac 1{\sqrt{-b}}\ln \left| 1-\frac a{2\sqrt{-b}}\right| .
\label{thfortytwo}
\end{equation}
In Ref. \cite{twentysix} another expression was used, true only when 
\begin{equation}
a^2+4b=\frac{121}9K^2\left( v^2-\frac{72}{121}\right) <0.
\label{thfortythree}
\end{equation}
This inequality, however, does not hold for $v$ close to $\sqrt{3/5}$.
Plugging Eq. (242) into Eq. (239) results in 
\begin{equation}
C_1=-\frac v{6\sqrt{2}}\left[ \ln \left( \frac v{6\sqrt{2}}+\sqrt{z_0}%
\right) -\ln \left| 1-\frac{11v}{6\sqrt{2}}\right| \right] +\sqrt{z_0}.
\label{thfortyfour}
\end{equation}
The pressure reads 
\begin{equation}
\kappa p=\frac{4C_2\sqrt{z}}{C_1+C_2\xi }-\frac 13\kappa \rho _0+7K^2x.
\label{thfortyfive}
\end{equation}
The condition $p\left( 0\right) >0$ is fulfilled when $C_1<2/11$. This holds
when $v$ varies near $\sqrt{3/5}=0.77.$ Thus, there are solutions with
positive pressure.

2) Case $d<0$. This means $\kappa \rho _1<11K^2$ and the limiting case $\rho
_1=0$ is possible. We shall discuss first the general case. Eqs.
(221),(222),(231),(235) hold. Conditions (229), (230) fix the constants 
\begin{equation}
C_1=\frac 12e^{\sqrt{-d}\xi _0}\left( \sqrt{z_0}-\frac{z_x\left( x_0\right) 
}{2\sqrt{-d}}\right) ,  \label{thfortysix}
\end{equation}
\begin{equation}
C_2=\frac 12e^{-\sqrt{-d}\xi _0}\left( \sqrt{z_0}+\frac{z_x\left( x_0\right) 
}{2\sqrt{-d}}\right) .  \label{thfortyseven}
\end{equation}
The pressure is given by Eq. (232) 
\begin{equation}
\kappa p=4\sqrt{-dz}\frac{C_2e^{\sqrt{-d}\xi }-C_1e^{-\sqrt{-d}\xi }}{C_2e^{%
\sqrt{-d}\xi }+C_1e^{-\sqrt{-d}\xi }}-\frac 13\kappa \rho _0+\frac 15\left(
3\kappa \rho _1+2K^2\right) x.  \label{thfortyeight}
\end{equation}
Positivity at the centre demands 
\begin{equation}
\frac{C_2-C_1}{C_2+C_1}>\frac{\kappa \rho _0}{12\sqrt{-d}}.
\label{thfortynine}
\end{equation}
A necessary condition is $C_2>C_1$, satisfied when 
\begin{equation}
z_x\left( x_0\right) =-\frac{\kappa \rho _0}3-\frac 25\left( K^2-\kappa \rho
_1\right) x_0  \label{thfifty}
\end{equation}
is positive, which leads to $\kappa \rho _1>6K^2$. Introducing the variable 
\begin{equation}
\Lambda =\frac{\kappa \rho _1}{K^2},  \label{thfiftyone}
\end{equation}
we obtain for its range $6<\Lambda <11$. Then $b<0$ and $\xi $ is given
again by Eq. (242). Eqs. (246), (247) become 
\begin{equation}
C_1=\frac 12\left( \sqrt{z_0}+\frac{z_x\left( x_0\right) }{2\sqrt{-b}}%
\right) ^{\sqrt{\frac db}}\left( \sqrt{z_0}-\frac{z_x\left( x_0\right) }{2%
\sqrt{-d}}\right) \left| 1-\frac a{2\sqrt{-b}}\right| ^{-\sqrt{\frac db}},
\label{thfiftytwo}
\end{equation}
\begin{equation}
C_2=\frac 12\left( \sqrt{z_0}+\frac{z_x\left( x_0\right) }{2\sqrt{-b}}%
\right) ^{-\sqrt{\frac db}}\left( \sqrt{z_0}+\frac{z_x\left( x_0\right) }{2%
\sqrt{-d}}\right) \left| 1-\frac a{2\sqrt{-b}}\right| ^{\sqrt{\frac db}}.
\label{thfiftythree}
\end{equation}
Obviously, $C_2>0$ and $C_2+C_1>0$, as required. Inequality (249) is still
very complicated. It becomes simpler if $d=b$ but this gives $\Lambda =3$
which is out of range. Another way to simplify it is to seek solution with $%
C_1=0$. Eq. (252) allows to rewrite this condition as 
\begin{equation}
K^2x_0^2=\frac{9\left( 11-\Lambda \right) }{27+9\Lambda -\Lambda ^2},
\label{thfiftyfour}
\end{equation}
while Eq. (249) becomes 
\begin{equation}
1>-\frac{\left( \kappa \rho _1\right) ^2x_0^2}{144d}.  \label{thfiftyfive}
\end{equation}
Plugging here the previous formula transforms Eq. (255) into 
\begin{equation}
\left( \Lambda +2\right) \left( \Lambda -6\right) <0,  \label{thfiftysix}
\end{equation}
satisfied when $-2<\Lambda <6$. This interval is adjacent to the allowed
interval for $\Lambda $, hence, this solution also possesses negative
pressure. We have been unable to find a positive pressure solution.

Let us discuss the subcase $\rho _1=0,$ i.e. $\rho =\rho _0$. Now Eq. (250)
shows that $z_x\left( x_0\right) <0$, therefore, $C_2<C_1$ and either Eq.
(249) is not satisfied or $y\left( 0\right) <0$, which is completely
unrealistic. The pressure is negative in this subcase. Eq. (222) holds as it
is, while Eq. (223) reads 
\begin{equation}
m=\left( \frac 16\kappa \rho _0+\frac 35K^2x_0\right) x_0^{3/2}.
\label{thfiftyseven}
\end{equation}
We have $b=K^2/5>0$ and consequently $a^2+4b>0$. Then Eq. (227) reads 
\begin{equation}
\xi =\frac 1{\sqrt{b}}\arcsin \frac{a+2bx}{\sqrt{a^2+4b}}-\frac 1{\sqrt{b}}%
\arcsin \frac a{\sqrt{a^2+4b}}.  \label{thfiftyeight}
\end{equation}
In Ref. \cite{twentysix} the expression 
\begin{equation}
\xi =\frac 1{\sqrt{-b}}arc\sinh \frac{a+2bx}{\sqrt{-a^2-4b}},
\label{thfiftynine}
\end{equation}
which holds only when $b<0$ and $a^2+4b<0$ was used in all cases and this is
incorrect. Formula (258) allows to make connection with the results of Ref. 
\cite{six}, where in addition $\rho _0=0$, so that the density vanishes and
the mass arises entirely from the electrostatic field energy. Eq. (250)
still shows that the pressure is negative. Eqs. (222) and (257) give the
ratio 
\begin{equation}
\frac{\left| e\right| }m=\frac 5{3\left| K\right| x_0}.  \label{thsixty}
\end{equation}
Now $a=0$ and 
\begin{equation}
z=1-\frac 15K^2x^2.  \label{thsixtyone}
\end{equation}
It is positive at the boundary when $\left| K\right| x_0<\sqrt{5}$ and this
means $\left| e\right| /m>\sqrt{5}/3$. The second term in Eq. (258) vanishes
and combining it with Eq. (235) yields 
\begin{equation}
y=C_1\exp \left( -\sqrt{\frac{11}2}\arcsin \sqrt{b}x\right) +C_2\exp \left( 
\sqrt{\frac{11}2}\arcsin \sqrt{b}x\right) .  \label{thsixtytwo}
\end{equation}
The passage to $\arccos $ scales $C_{1,2}$ and interchanges their places,
giving Eq. (4.6) from Ref. \cite{six}.

3) Case $d>0$ . Now we have $\kappa \rho _1>11K^2$ which permits the limit $%
K=0$. The two cases do not differ essentially. Eq. (236) gives $y$. The
constants read 
\begin{equation}
C_1=\sqrt{z_0}\sin \sqrt{d}\xi _0+\frac{z_x\left( x_0\right) }{2\sqrt{d}}%
\cos \sqrt{d}\xi _0,  \label{thsixtythree}
\end{equation}
\begin{equation}
C_2=\sqrt{z_0}\cos \sqrt{d}\xi _0-\frac{z_x\left( x_0\right) }{2\sqrt{d}}%
\sin \sqrt{d}\xi _0.  \label{thsixtyfour}
\end{equation}
The pressure is given by a formula, similar to Eq. (248). Positivity at the
centre is ensured by 
\begin{equation}
\frac{C_1}{C_2}>\frac{\kappa \rho _0}{12\sqrt{d}}.  \label{thsixtyfive}
\end{equation}
The value of $z_x\left( x_0\right) $ is positive, while $b<0$. Hence, $\xi $
is given by Eq. (242), like in Ref. \cite{twentyfive}. Condition (265) is
even more complicated than Eq. (249) and simplifications do not seem
possible. This is true even for the uncharged case, discussed in the
mentioned reference. We shall remark only that the simple case $\kappa \rho
_1=\kappa \rho _0=x_0=1$, $K=0$ has positive pressure profile. For small
enough $K$, the charged case must behave the same way. This ends the
discussion of the case $l=3$.

Let us study next the case $l=2$ when $z$ is given by Eq. (220). Eq. (224)
is not soluble unless $b\equiv -\kappa \rho _1/5=0$ meaning that the density
is constant. Then Eq. (224) becomes a particular case of the hypergeometric
equation 
\begin{equation}
\chi \left( \chi -1\right) y_{\chi \chi }+\left[ \left( \eta _1+\eta
_2+1\right) \chi -\eta _3\right] y_\chi +\eta _1\eta _2y=0,
\label{thsixtysix}
\end{equation}
where $\chi =ax$, $\eta _1+\eta _2+1=1/2$, $\eta _3=0$, $\eta _1\eta
_2=K^2/2a$. These relations give 
\begin{equation}
\eta _1=-\frac 14\pm \frac 14\sqrt{1-\frac{8K^2}a}  \label{thsixtyseven}
\end{equation}
and $\eta _{1,2}\in [-1/2,0)$. The reality of $\eta _1$ is ensured by 
\begin{equation}
\kappa \rho _0>23K^2.  \label{thsixtyeight}
\end{equation}
Thus, in the charged case it is not possible to put $\rho =\rho _0=0$. This
case was studied by Wilson \cite{twentythree} who did not recognize the
appearance of the hypergeometric function. He assumed that the pressure was
positive and developed a series expansion for $y$. Now, since $\eta _3=0$,
the two fundamental solutions of Eq. (266) are \cite{fiftynine} 
\begin{equation}
y_1=\chi F\left( \eta _1+1,\frac 12-\eta _1,2;\chi \right) ,
\label{thsixtynine}
\end{equation}
\begin{equation}
y_2=F\left( \eta _1,-\frac 12-\eta _1,\frac 12,1-\chi \right) .
\label{thseventy}
\end{equation}
This is because $1+\eta _1+\eta _2$, $\pm \eta _{1,2}$ are not integers. The
first solution is unphysical since $e^\nu $ vanishes at $r=0$. The second
has a finite limit 
\begin{equation}
y_2\left( 0\right) =\frac{\sqrt{\pi }}{\Gamma \left( \frac 12-\eta _1\right)
\Gamma \left( 1+\eta _1\right) },  \label{thseventyone}
\end{equation}
so a linear combination between the two solutions is a candidate for a
regular metric.

In Sec. VI we have shown that when $e^\nu $ is given by the simple ansatz
(145) with $k=1$, $z$ is expressed through hypergeometric functions. Here we
see that, in a rather symmetrical way, when $z$ is given by the similar
ansatz (50) (or (220) with $\rho _1=0$), $y$ becomes a hypergeometric
function. In this paper the emphasis is laid on solutions in elementary
functions, so we shall not pursue this issue further, except in the case
where $K=0$. Then Eq. (266) becomes 
\begin{equation}
\left( \chi -1\right) y_{\chi \chi }+\frac 12y_\chi =0  \label{thseventytwo}
\end{equation}
and is easily solved 
\begin{equation}
e^\nu =\left[ 2\left( 1-ar^2\right) ^{1/2}+a_1\right] ^2.
\label{thseventythree}
\end{equation}
This is the expected Schwarzschild interior solution \cite{fiftyeight},
contrary to the claims in Ref. \cite{twentythree}.

A model for a superdense star with the ansatz 
\begin{equation}
z=\frac{1-a_2r^2}{1+a_1a_2r^2}  \label{thseventfour}
\end{equation}
has been discussed in the uncharged case for $a_1=2$ \cite{sixty,sixtyone}, $%
a_1=7$ \cite{sixtytwo} and a set of discrete values for $a_1$ \cite
{fortythree}. Recently, it was studied for arbitrary $a_1$ both in the
uncharged \cite{fortyfour} and the charged case \cite{fortyfive}. The charge
function is chosen to be 
\begin{equation}
q=\frac{Ka_2r^3}{1+a_1a_2r^2},  \label{thseventyfive}
\end{equation}
which simplifies considerably the equation for $y$. Introducing the variable 
\begin{equation}
\eta =\left( \frac{a_1}{a_1+1}\right) ^{1/2}\left( 1-a_2r^2\right) ^{1/2},
\label{thseventysix}
\end{equation}
one obtains 
\begin{equation}
\left( 1-\eta ^2\right) y_{\eta \eta \eta }-\eta y_{\eta \eta }+d_1y_\eta =0,
\label{thseventyseven}
\end{equation}
\begin{equation}
d_1=a_1+2-\frac{2K^2}{a_1},  \label{thseventyeight}
\end{equation}
where $d_1>0$ is required.

Formally, Eq. (277) is a subcase of Eq. (224) with $a=0$, $b=1$, $%
x\rightarrow r$, $y\rightarrow y_\eta $, so we can use the machinery
developed there. Eq. (227) gives $\xi =\arcsin \eta $. We may change the
variable to $\delta =\arccos \eta $ because $\xi +\delta =\pi /2$ and Eq.
(226) is invariant under this change. Hence, $y_\eta $ is given by Eq. (236)
and $y$ may be found by integration. Use of trigonometric equalities helps
to find an expression with two terms 
\begin{equation}
y=a_4\left\{ \frac{\cos \left[ \left( \sqrt{d_1}+1\right) \delta +a_3\right] 
}{\sqrt{d_1}+1}-\frac{\cos \left[ \left( \sqrt{d_1}-1\right) \delta
+a_3\right] }{\sqrt{d_1}-1}\right\} ,  \label{thseventynine}
\end{equation}
where the notation $C_1=-2a_4\sin a_3$, $C_2=2a_4\cos a_3$ has been used.
This formula was found in Refs. \cite{fortyfour,fortyfive} by resorting
first to Gegenbauer functions and then to Chebyshev polynomials. The
peculiarities of Eq. (224) permit a straightforward derivation in elementary
functions.

Finally, let us discuss a solution of Nduka \cite{thirtyfive}, belonging to
the class $\left( \lambda ,q\right) $. He chose a constant $z\neq 1$ and
generalized the uncharged solution of Ref. \cite{thirtyfour} by taking $l=1$%
, i.e. $q=Kr$. Then Eq. (24) turns into the Euler equation. Repeating the
analysis referring to Ref. \cite{thirtyseven}, after Eq. (157), we see that $%
y$ is singular at $r=0$. The same is true for the density, $\rho \sim 1/r^2$.

\section{The case $\left( \lambda ,n\right) $}

In this case the fluid satisfies the linear equation of state (17) and some
ansatz for $\lambda $ is given. The other metric component is found from Eq.
(27) - a linear second-order equation for $y$.

Let us study first the case $z=c<1$ which leads to singular pressure and
density, but provides the opportunity to generalize the well-known KT
solution \cite{fortyone,fortytwo,fortysix,sixtythree,sixtyfour}. Eq. (27)
becomes the Euler equation when $p_0=0$ ($\gamma $-law) 
\begin{equation}
r^2y^{\prime \prime }+k_0ry^{\prime }-k_1y=0,  \label{theighty}
\end{equation}
\begin{equation}
k_0=\frac{3-n}{n+1}\text{,\qquad }k_1=\frac{1-c}c,  \label{theightyone}
\end{equation}
whose solution is 
\begin{equation}
e^\nu =C_1r^{1-k_0+2k_2},  \label{theightytwo}
\end{equation}
\begin{equation}
k_2=\frac 12\left[ \left( 1-k_0\right) ^2+4k_1\right] ^{1/2}.
\label{theightythree}
\end{equation}
It can be checked that $1-k_0+2k_2>0$. We have included only the
non-divergent at $r=0$ solution. The density and the charge function are
given by Eqs. (166), (168) respectively 
\begin{equation}
\left( n+1\right) \kappa \rho =\frac{\left( 1-k_0+2k_2\right) c}{r^2},
\label{theightyfour}
\end{equation}
\begin{equation}
\frac{q^2}{r^2}=\left( 1-c\right) \left[ 1-\frac{4n+\left( n+1\right) ^2}{%
\left( n+1\right) ^2}c\right] .  \label{theightyfive}
\end{equation}
The density is positive and has a pole at the centre. Obviously 
\begin{equation}
c\leq \frac{\left( n+1\right) ^2}{4n+\left( n+1\right) ^2}
\label{theightysix}
\end{equation}
should hold. This solution is a charged generalization of the KT solution
which is recovered when the equality holds. Due to $p_0=0$ it does not have
a boundary and is not asymptotically flat. Another generalization was
offered in Ref. \cite{forty} but it alters many of its properties, including
the equation of state. If $p_0\neq 0$ Eq. (27) becomes 
\begin{equation}
r^2y^{\prime \prime }+k_0ry^{\prime }+\left( -k_1+k_3r^2\right) y=0,
\label{theightyseven}
\end{equation}
\begin{equation}
k_3=\frac{2\kappa p_0}{\left( n+1\right) c}.  \label{theightyeight}
\end{equation}
Its solution is given by Bessel functions \cite{fortynine} 
\begin{equation}
y=r^{\frac{1-k_0}2}\left[ C_1J_{k_2}\left( \sqrt{k_3}r\right)
+C_2Y_{k_2}\left( \sqrt{k_3}r\right) \right] .  \label{theightynine}
\end{equation}

Let us next discuss the ansatz (50), imposed by the condition of constant $%
T_0^0$, like in Sec. III. Introducing $x=ar^2$ turns Eq. (27) into 
\begin{equation}
x\left( x-1\right) y_{xx}+\left[ \frac{n+5}{2\left( n+1\right) }x-\frac 2{n+1%
}\right] y_x+Dy=0,  \label{thninety}
\end{equation}
\begin{equation}
D=\frac 1{2\left( n+1\right) }\left( 3n+1-\frac{\kappa p_0}a\right) .
\label{thninetyone}
\end{equation}
This is once again the hypergeometric equation and one of its fundamental
solutions is $y_1=F\left( \eta _1,\eta _2,\eta _3;x\right) $ where 
\begin{equation}
\eta _2=\varepsilon -\eta _1,  \label{thninetytwo}
\end{equation}
\begin{equation}
\eta _3=\frac 2{n+1},  \label{thninetythree}
\end{equation}
\begin{equation}
\eta _1\eta _2=D,  \label{thninetyfour}
\end{equation}
\begin{equation}
\varepsilon =\frac{3-n}{2\left( n+1\right) }.  \label{thninetyfive}
\end{equation}
The solution is determined by the parameter $n$ and the ratio $\kappa p_0/a$
which enters the definition of $D$. Eq. (294) shows that we can take instead 
$\eta _1$ as a second independent parameter. Eq. (291) provides one upper
limit for $D$: $D\leq \frac{3n+1}{2\left( n+1\right) }$. In the physical
range $0\leq n\leq 1$ it increases from $1/2$ to $1$. Eq. (294) is a
quadratic equation for $\eta _1$ and it is real when $D\leq \left(
3-n\right) ^2/16\left( n+1\right) ^2$. This second limit decreases when $n$
increases, from $9/16$ to $1/16$. There is a value $n_0=0.026$ where both
limits meet. Thus, the first inequality is stronger when $n\in \left[
0,n_0\right] $, while the second dominates for $n\in \left[ n_0,1\right] $.
The interval $\left[ 0,n_0\right] $ is also the range of $n$ where $\eta _1$
is real when $p_0=0$. The case $n=1$ is degenerate because $\eta _3=1$. This
affects the second fundamental solution \cite{fiftynine} (Vol.1). When $n<1$
it is 
\begin{equation}
y_2=x^{\frac{n-1}{n+1}}F\left( \eta _1+\frac{n-1}{n+1},\frac 12-\eta _1,%
\frac{2n}{n+1};x\right)  \label{thninetysix}
\end{equation}
and has a pole at $r=0$. Therefore, we shall discard it. When $n=1$ and $%
\eta _1,\frac 12+\eta _1$ are not integers, we have 
\begin{equation}
y_2=F\left( \eta _1,\frac 12-\eta _1,\frac 12;1-x\right) .
\label{thninetyseven}
\end{equation}
This function diverges at $x=0$ due to the formula 
\begin{equation}
F\left( \eta _1,\eta _2,\eta _3;1\right) =\frac{\Gamma \left( \eta _3\right)
\Gamma \left( \eta _3-\eta _2-\eta _1\right) }{\Gamma \left( \eta _3-\eta
_1\right) \Gamma \left( \eta _3-\eta _2\right) }.  \label{thninetyeight}
\end{equation}
When $\eta _1=-k$ or $-k+\frac 12$, $k$ being a positive integer, $y_2$ is
given by the so-called Kummer series and is reduced essentially to a
polynomial. When $\eta _1=k$ or $k+\frac 12$ the same is true, due to the
symmetry of $F$ with respect to its first two arguments. In the following we
shall use mainly $y_1$ which is well-defined by the hypergeometric series
and converges for $x\in \left[ 0,1\right] $. The generic case may be divided
into three subcases: 1) $D=0$, 2) $D<0$ and 3) $D>0$. We shall search them
in turn for elementary solutions.

1) $D=0$. Then $\eta _1=0$, $\eta _2=\varepsilon $, $\kappa p_0/a=3n+1$ and $%
y$ is expressed through the incomplete $\beta $-function 
\begin{equation}
y=C_1+C_2\int x^{-\frac 2{n+1}}\left( 1-x\right) ^{-1/2}dx.
\label{thninetynine}
\end{equation}
When $n=1$ the integral is elementary and reads 
\begin{equation}
\ln \frac{1-\left( 1-x\right) ^{1/2}}{1+\left( 1-x\right) ^{1/2}}.
\label{thh}
\end{equation}
The logarithmic singularity at the beginning is obvious. When $n<1$ it
becomes a pole so that $C_2=0$ is required. Then, however, $q=0$ and the
solution is uncharged. It is easy to find that $\kappa p=-a$ and $\rho +3p=0$%
. We obtain ESU \cite{fifty,fiftytwo} and Eq. (299) represents its
ill-defined generalization to the charged case.

2) $D<0$. Ambiguity is fixed by accepting that $\eta _1<0$, $\eta _2>0$. A
rather special subcase is $\eta _1=-1/2$. Then $\eta _2=\eta _3$ and 
\begin{equation}
e^\nu =1-x=z.  \label{thhone}
\end{equation}
Eq. (10) gives $\rho +p=0$ while $q=0$, $\rho =\frac{p_0}{n+1}$. Thus for
any $n$ we obtain the de Sitter metric whose charged generalizations were
discussed in Sec. IV. This effective change in the equation of state was
necessary because Eq. (27) is undefined for $n=-1$, and is possible when $%
\rho $ and $p$ are constant. It was noticed in the uncharged case for ESU in
Ref. \cite{fortyseven}.

Another interesting elementary subcase is given by $\eta _1=-k$ when $F$
becomes a polynomial obtained from the first $k+1$ terms of the
hypergeometric series. It may be written also as 
\begin{equation}
F\left( -k,\varepsilon +k,\eta _3;x\right) =\frac{k!}{\left( \eta _3\right)
_k}P_k^{\left( \eta _3-1,-1/2\right) }\left( 1-2x\right) =\frac{\left(
1/2\right) _kk!}{\left( \eta _3\right) _k\left( \varepsilon \right) _k}%
C_{2k}^{\left( \varepsilon \right) }\left( \sqrt{1-x}\right) ,
\label{thhtwo}
\end{equation}
where $\left( a\right) _k=a\left( a+1\right) ...\left( a+k-1\right) ,$ $P$
and $C$ are respectively Jacobi and Gegenbauer orthogonal polynomials \cite
{fiftynine} (Vol.2). Unfortunately, this nice solution is plagued, like the
de Sitter solution, and like all solutions with negative $D$, by tension. To
show this let us plug our ansatz in Eqs. (166)-(168) and replace $\kappa
p_0/a$ by $D$ through Eq. (291): 
\begin{equation}
\frac{\left( n+1\right) q^2}{ar^4}=2\left( n+1\right) D-2\left( 1-x\right)
\nu _x,  \label{thhthree}
\end{equation}
\begin{equation}
\frac{\left( n+1\right) \kappa p}a=-\left( n+1\right) +2\left( n+1\right)
D+2n\left( 1-x\right) \nu _x.  \label{thhfour}
\end{equation}
Since $D<0$, $q^2>0$ only if $\nu _x<0$. But then all three terms in the
pressure are negative.

3) $D>0$. Then only the combination $\eta _{1,2}>0$ is possible, for when $%
\eta _1<0$, $\eta _2>0$. Eqs. (292), (294) yield $0<\eta _1<\varepsilon $.
The right limit decreases with $n$ and always $\varepsilon \leq 3/2$. An
interesting subcase is $n=1/3$ . We have $\varepsilon =1$, $\eta _3=1/2$, $%
D=\eta _1\left( 1-\eta _1\right) $. Then $y_1$ degenerates into 
\begin{equation}
y=C_1\frac{\sin \left( 1-2\eta _1\right) R}{\left( 1-\eta _1\right) \sin 2R},
\label{thhfive}
\end{equation}
\begin{equation}
x=\sin ^2R.  \label{thhsix}
\end{equation}
The ranges are $0<\eta _1<1$, $0<D<1/4$, $4/3<\kappa p_0/a<2$. When $\eta _1$
becomes bigger than $1/2$ the sign of $C_1$ should be changed to keep $y$
positive. Eqs. (303), (304) become 
\begin{equation}
\frac{q^2}{3ar^4}=\frac 23D-\frac 12\left( 1-x\right) \nu _x,
\label{thhseven}
\end{equation}
\begin{equation}
\frac{\kappa p}a=\frac 12\left( 1-x\right) \nu _x-1+2D.  \label{thheight}
\end{equation}
We demand the positivity of $p$. This means at least 
\begin{equation}
\frac 12\left( 1-x\right) \nu _x>\frac 12,  \label{thhnine}
\end{equation}
i.e. $\nu _x$ must be positive. Then, however, the first term in Eq. (307)
cannot compensate the second because $2D/3<1/6$. If we make $q^2$ positive,
the pressure becomes negative in complete analogy with the case $D<0$.

Finally, let us point out that Eq. (27) may be obtained by linearization of
Eq. (33). Let us search for simple solutions of this equation. In Sec. IV we
solved the case $n=-1$, when $Y=0$ using Eq. (30). Let us assume now that 
\begin{equation}
\frac 2{r^4}\left( r^2M^{\prime }\right) ^{\prime }=\frac{4\kappa p_0}{%
\left( n+1\right) r}.  \label{thhten}
\end{equation}
Quite interestingly, this again leads to the ansatz (50) 
\begin{equation}
z=1-\frac{\kappa p_0}{3\left( n+1\right) }r^2  \label{thheleven}
\end{equation}
and $\kappa p_0/a=3\left( n+1\right) $, $D=-\frac 1{n+1}$. Then Eqs. (292),
(294) give 
\begin{equation}
\eta _1=\frac{3-n\pm \left( n+5\right) }{4\left( n+1\right) },
\label{thhtwelve}
\end{equation}
so that $\eta _1$ and $n$ are not independent. The first root leads to $\eta
_1=\eta _3$ and Eq. (301). This is the trivial $Y=0$ solution of Eq. (33).
The other root gives $\eta _1<0$ and $D<0$, leading to negative pressure.
Something more, from the general solution of Eq. (33) 
\begin{equation}
Y=\frac 1{z^{1/2}r^{\frac 4{n+1}}\left( C+\frac 12\int z^{-3/2}r^{\frac{n-3}{%
n+1}}dr\right) },  \label{thhthirteen}
\end{equation}
it is seen that $Y$ (and the pressure) has a pole at the centre, even when $%
C=0$. In conclusion, we have been unable to find realistic solutions with
positive pressure for the ansatz (50).

\section{Discussion and conclusions}

Charged static perfect fluids have attracted considerably less attention
than the uncharged ones, the number of papers being roughly an order of
magnitude smaller. Solutions have been rare and the introduction hints how
diverse were the approaches to their study. Charged dust occupies the first
place in popularity and puts forth the idea of point-like particles. The
rest is a mixture of generalizations: of the Schwarzschild idea about
incompressibility, which may mean constant $T_0^0$ or constant density in
the charged case, with the limiting case of vanishing density and
electromagnetic mass models, generalization of the idea about vacuum
polarization which brings forth the de Sitter solution, generalization
simply of well-known uncharged solutions like those of Adler, Kuchowicz,
Klein, Mehra, Vaidya-Tikekar and others, generalization of Weyl type
connections, well studied in the electrovac case.

In this paper we have tried to show that the charged case has a life of its
own when subjected to a natural and organized classification scheme.
Surprisingly, in many respects it looks simpler than the uncharged case. The
presence of the charge function serves as a safety valve, which absorbs much
of the fine tuning, necessary in the uncharged case. The general formulae
derived here show that the abundance of elementary solutions is probably
bigger than in the traditional case. The proposed scheme becomes rather
trivial there, representing ansatze mainly for $\lambda $, $\nu $ and
sometimes $\rho $. In the charged case, however, it allows to sort out the
different ideas mentioned above. Thus constant $T_0^0$ leads to the cases $%
\left( \lambda =1-ar^2,*\right) $, while constant density - to the much more
difficult cases $\left( \nu ,\rho \right) $ and $\left( \lambda ,\rho
\right) $, similar to $\left( \nu ,n\right) $ and $\left( \lambda ,n\right) $%
. Electromagnetic mass models are subcases, often spoiled by negative
pressure. The point-like idea seems to be viable only for CD, leading to
flat spacetime when the pressure does not vanish. The generalization of
uncharged solutions firmly occupies the cases $\left( \lambda ,q\right) $
and $\left( \nu ,q\right) $. Models with $n=-1$ are intimately related to
the soluble case $n\neq -1$ and are much richer than their traditional
protagonist - the de Sitter solution. The other models with negative
pressure, which seem worth being studied, are the generalizations of ESU
with $n=-1/3$. Finally, the Weyl type relations, so successful in electrovac
and CD environments, seem rather out of place in a 'pressurized' perfect
fluid.

Another advantage of this classification is that it delineates the degree of
difficulty and the most convenient points for attack of the problem. The
easiest cases seem to be $\left( \lambda ,Y\right) $ and $\left( \nu
,Y\right) $. If we insist on authentic fluid characteristics and not their
combinations, then $\left( \nu ,p\right) $ and $\left( \nu ,q\right) $ are
the best candidates. On the other hand, the most difficult are $\left( \rho
,p\right) $ and $\left( p,q\right) $, which seem to be accessible only
numerically. The most unpredictable case is probably $\left( \lambda ,\nu
\right) $ since there is no control on exactly those characteristics which
must satisfy the majority of regularity and positivity conditions.

A different division is between general and special cases. As was noticed in
the introduction, a known solution belongs to any of the general cases; a
solution of $\left( \nu ,q\right) $ may be written as $\left( \lambda
,p\right) $ or even as $\left( \rho ,p\right) $ solution. Only the
simplicity of two of the five functions $\nu ,\lambda ,\rho ,p,q$ betrays
where was the starting point. The special cases we discussed, apart from the
ansatz (50), are essentially three: $\rho +p=0$ and $\left( \lambda
,n\right) $, $\left( \nu ,n\right) $. All of them have a linear equation of
state. It should be mentioned that all integrable uncharged solutions with $%
\gamma $-law found in Ref. \cite{fortysix} have their charged
generalizations with $p_0\neq 0$ and, hence, have a boundary. The case $n=0$%
, $p_0=0$ represents CD and arises as an electrification of the trivial
uncharged dust solution (flat spacetime) as seen from Eqs. (95)-(98). Its
boundary may be put anywhere, since $p=0$ everywhere. The case $n=-1$
generalizes the de Sitter solution into a bunch of new solutions,
characterized by the mass function, subjected to several mild restrictions.
It was completely solved in Sec. IV. A second, this time ill-defined
generalization, is given by Eq. (313). The de Sitter solution appears also
as a special case in Sec. X (the class $D<0$ with $\eta _1=-1/2$). Since Eq.
(33) excludes the case $n=-1$ this becomes possible thanks to an effective
change in the equation of state. The option $n=-1/3$ in the uncharged case
is a mark of ESU. One possible generalization of it was performed in Sec.
VII (see Eqs. (176), (177) and their discussion). The ESU appears also as
the special class $D=0$ in Sec. X. The case $n=-1/5$ is special in the
charged theory too, but its metric is singular, as was found in Eqs. (202),
(203). The KT solution is generalized in Eqs. (282)-(286). In the uncharged
case these exhaust the integrable cases of a complicated Abel differential
equation of the second kind. In the charged case it is replaced by a linear
first-order Eq. (19), which has many other solutions, discussed in Sec. VII,
or by a linear second-order Eq. (27), discussed in Sec. X.

From a mathematical viewpoint, the charged case delivers a surprising
variety of equations like those of Euler (157), (280), Bernoulli (75),
(313), Riccati (30)-(33), Emden-Fowler (113), Abel (37), (102), (114) and
the hypergeometric equation (266), (290). The cases containing $\lambda $
quickly lead to special functions whose spectrum is also rich: $%
\mathop{\rm Ei}
$, $cn$, Bessel, hypergeometric, incomplete $\beta $-functions, Jacobi,
Gegenbauer and Chebyshev polynomials, etc. Therefore, a more thorough study
would require computer simulations of special functions, which are easier
than purely numeric simulations.

From a physical viewpoint, the most interesting results are, in our opinion,
the following:

1) The case of constant $T_0^0$ is solved by a simple algorithm involving
algebraic operations, one differentiation and one simple integration upon a
generating function $Y$, which satisfies few simple inequalities, Eqs.
(51)-(53).

2) The charged de Sitter case is completely soluble in terms of the mass or
the charge function, Eqs. (56)-(59). An example is given by Eqs. (71)-(74).
There is a generalization with positive pressure in Eqs. (76)-(78). An
example is given by Eqs. (79)-(82).

3) The general solution for a given pressure may be written as three
contributions to $z$; from regular CD, from general CD and from the
pressure, see Eq. (83). There is a general CD solution which creates a halo
around charged fluid balls and may postpone their junction to a RN solution.
An example is given by Eqs. (115)-(140).

4) The solutions of Korkina-Durgapal possess realistic charge
generalizations given by Eqs. (142)-(145). An example is given by Eqs.
(148)-(153).

5) Fluids with linear equation of state are integrable for any $n$ as seen
from Eqs. (19), (27), (33), (159)-(165). Elementary solutions may be found
by several simplification techniques. Physically realistic is an infinite
series of models given by Eqs. (173), (174), (178)-(180). Realistic examples
are given by Eqs. (182), (185). The status of the model given by Eq. (190)
is still unknown.

6) The case $\left( \nu ,\rho \right) $ is closely connected to $\left( \nu
,n\right) $ and is much more difficult than its companion with constant $%
T_0^0$, even when the density is constant or zero. An example with special
functions is given by Eq. (213), another one - by Eq. (216). Their physical
status is still undetermined.

7) The $\left( \rho ,q\right) $ case discussed in Ref. \cite{twentysix} is
incomplete and has errors. It is treated in detail in Sec. IX and its
connection with Ref. \cite{six} is elucidated. It has several realistic
cases, but electromagnetic mass models all seem to have negative pressure.
Realistic examples are given by Eqs. (218), (234), (238)-(245) and Eqs.
(263)-(265) with the values specified after them. A formal, but intriguing
parallel is drawn to star models with simple spatial geometry \cite
{fortyfive}.

8) The solution of Wilson \cite{twentythree} is expressed by a
hypergeometric function and reduces to the Schwarzschild interior solution
when the charge vanishes. Its realistic status is not known.

9) The KT solution has a charged generalization which is likewise singular,
but may serve as a focal point to some regular solution, see Eqs.
(282)-(286).

10) Solutions of the case $\left( \lambda ,n\right) $ with constant $T_0^0$
are expressed in hypergeometric functions, see Eqs. (290), (291). All
degenerate elementary solutions have either negative pressure or
singularities.

\end{document}